 \documentclass[final,5p,times,twocolumn]{elsarticle}

\usepackage{amssymb}
\usepackage{lipsum}
\usepackage{xcolor}
\usepackage{tabularx}
\usepackage{booktabs}
\usepackage{geometry}
\usepackage{float}
\usepackage{hyperref}

\usepackage{xcolor}
\usepackage{pgfgantt}
\usepackage{multirow}
\usepackage{color, colortbl}
\definecolor{Gray}{gray}{0.9}
\definecolor{LightCyan}{rgb}{0.88,1,1}
\usepackage{makecell}
\usepackage{caption}
\usepackage[export]{adjustbox}

\usepackage{listings}

\journal{Journal of Information Security and Applications}

\begin{document}

\begin{frontmatter}

\title{NeuroIDBench: An Open-Source Benchmark Framework for the Standardization of Methodology in Brainwave-based Authentication Research}

\author[a]{Avinash Kumar Chaurasia}

\author[b]{Matin Fallahi}

\author[b]{Thorsten Strufe}

\author[a]{Philipp Terhörst}

\author[a,b]{Patricia Arias Cabarcos}

\affiliation[a]{organization={University of Paderborn},
            city={Paderborn},
            country={Germany}}
\affiliation[b]{organization={KASTEL Security Research Labs - KIT},
            city={Karlsruhe},
            country={Germany}}

\begin{abstract}

Biometric systems based on brain activity have been proposed as an alternative to passwords or to complement current authentication techniques. By leveraging the unique brainwave patterns of individuals, these systems offer the possibility of creating authentication solutions that are resistant to theft, hands-free, accessible, and potentially even revocable. However, despite the growing stream of research in this area, faster advance is hindered by reproducibility problems. Issues such as the lack of standard reporting schemes for performance results and system configuration, or the absence of common evaluation benchmarks, make comparability and proper assessment of different biometric solutions challenging. Further, barriers are erected to future work when, as so often, source code is not published open access. To bridge this gap, we introduce NeuroIDBench, a flexible open source tool to benchmark brainwave-based authentication models. It incorporates nine diverse datasets, implements a comprehensive set of pre-processing parameters and machine learning algorithms, enables testing under two common adversary models (known vs unknown attacker), and allows researchers to generate full performance reports and visualizations. We use NeuroIDBench to investigate the shallow classifiers and deep learning-based approaches proposed in the literature, and to test robustness across multiple sessions. We observe a 37.6\% reduction in Equal Error Rate (EER) for unknown attacker scenarios (typically not tested in the literature), and we highlight the importance of session variability to brainwave authentication. All in all, our results demonstrate the viability and relevance of NeuroIDBench in streamlining fair comparisons of algorithms, thereby furthering the advancement of brainwave-based authentication through robust methodological practices. 
\end{abstract}


\end{frontmatter}

\section{Introduction}
\label{sec:introduction}
In today's rapidly changing security landscape, the authentication process is a crucial element of access control \cite{mohamed2022systematic}. The increasing impracticality and safety concerns associated with conventional authentication methods, such as passwords \cite{alroomi2023measuring,taneski2014password}, are driving the development of novel biometric solutions. In this domain, user recognition based on brain activity has gained increasing attention \cite{gui2019survey}, especially with advancements in consumer-grade EEG (Electroencephalogram) technology \cite{emotivEPOCChannel, Neurable, Muse, Varjo}. 
 A notable application scenario is Extended Reality (XR), where seamless and secure biometric authentication systems are urgently needed \cite{stephenson2022sok}. Besides, brain biometrics bring broader benefits, such as enabling hands-free authentication, resistance to observation, inherent liveness detection, broad universality, and potential for revocability \cite{gui2019survey,zhang2021review,lin2018brain}. 
 
Unfortunately, the research in brainwave-based authentication is beset by problems of generalizability, comparability, and reproducibility. To begin, dataset sharing is sharply curtailed by ethical considerations and privacy laws. Therefore, many studies are based on self-collected undisclosed datasets, which often leads to results that are potentially overfitted or optimized to the specific characteristics of those datasets, with uncertain applicability to other, independent datasets. Also, the size of the datasets used in studies is severely limited by the complexity and resource intensiveness of broad data collection. In fact, most brainwave-based authentication studies have much fewer than 50 subjects  \cite{gui2019survey,bidgoly2020survey}, so that the evaluations are normally conducted on very small datasets. Consequently, the models and hyperparameters are prone to overfitting, which means the reported results will likely fail to generalize. To make matters worse, too many authors either neglect making available their source code or provide code which is inoperable, with the consequence that researchers are prevented from reproducing and building upon previous work. Clearly, under circumstances such as these, our subcommunity of authentication researchers will not make any significant progress. That is why the time has come for us to consolidate our practice and advance the research toward consistent, reproducible, and generalizable findings. To address this gap, we make the following contributions:

\begin{itemize}
    \item [] (1) We present \textbf{NeuroIDBench}, \textbf{the first comprehensive open-source\footnote{\url{https://github.com/Avichaurasia/NeuroIDBench.git}} benchmark tool designed to assist researchers in evaluating their brainwave-based authentication approaches}. NeuroIDBench is engineered to allow for the easy and flexible integration of new datasets and methods, enabling fair comparability. NeuroIDBench incorporates nine public datasets, which together contain a substantial number of subjects (n = 285, Table \ref{tab:erp_datasets}) and multi-session recordings. To determine the impact of preprocessing on performance, NeuroIDBench allows to apply different sample rejection thresholds as well as sample length parameters~(Sec \ref{sec:pre}) and feature extraction methods (Sec \ref{sec:feature}). Also, it implements a baseline with the most popular authentication algorithms in the state of the art, including shallow classifiers and deep learning-based approaches.  Lastly, it incorporates two attack scenarios under which to test the algorithms \cite{mansfield2002best}: i) \textit{known attacker}, where the attacker is known to the authentication model (i.e., enrolled in the system), and, ii) unknown attacker, a more realistic situation in which the attacker biometric data has not been seen by the system previous to the attack.  
    
\item [] (2) \textbf{We use NeuroIDBench to build a comprehensive benchmark of popular brainwave-based authentication algorithms, providing the first unbiased comparison of approaches}. Namely, we compare six shallow classifiers: SVM (Support Vector Machine), Random Forest, KNN (K-Nearest Neighbors), LDA (Linear Discriminant Analysis), Naive Bayes (NB), and Logistic Regression (LR). Besides this benchmark, which covers the biggest share of solutions in the literature, we provide a second benchmark for the recent stream of work on deep learning-based approaches. Specifically, NeuroIDBench implements the representative similarity-based technique using Twin Neural Networks (Sec. \ref{sec:shallow}).  Beyond performance across an extensive list of publicly available EEG datasets, we focus our comparative analysis on two under-explored aspects in current research:\textit{ How robust are different solutions in multi-session scenarios?} (Sec. \ref{sec:session}) and \textit{What is the impact of the adversary model? }(Sec. \ref{sec:seen})

\end{itemize}

Among the key findings in the analysis we observe that Random Forest consistently outperforms other classifiers and its performance is comparable to that of Twin Neural Networks. As the latter brings not only better performance but scalability advantages, we recommend more research on deep learning-based approaches to move forward, which is also the path followed in other biometric communities \cite{wang2021deep,rim2021fingerprint,minaee2023biometrics}. Furthermore, our results confirm that integrating Power Spectral Density (PSD) with an Autoregressive (AR) model of order 1 stands out as the most effective feature extraction approach. When investigating the impact of the adversary model, our results indicate that the average Equal Error Rate (EER) for known attackers is 2.87\%, while for unknown attackers, it is 4.6\%.  
 This trend in performance degradation underscores the need to evaluate brainwave authentication solutions under unknown attacker scenarios to provide a realistic performance indicator, a practice that is uncommon in current work. Variations in the performance across datasets confirm that brainwave-based authentication algorithms validated on single datasets are not representative, highlighting the need for more robust validation in the community. Finally, the observed disparities in the results for single-session versus multi-session authentication reveal a substantial gap that requires future research in this area.

\section{Background and Related Work}
\label{sec:background}

This section lays out the main terminology and background information on brainwave-based biometrics and reviews related work in the area, discussing reproducibility.

\subsection{Using brainwaves for Biometric Authentication}

\textbf{Brainwaves}, or EEG signals, represent the electrical activity produced by neuron interactions in the brain~\cite{niedermeyer2005electroencephalography,hu2019eeg}. These signals are commonly recorded from the scalp using specialized sensors. Originally developed for medical purposes~\cite{haas2003hans}, such as diagnosing neurological disorders \cite{adeli2010automated}, EEG technology has expanded its applications significantly. Its usage now extends to Brain-Computer Interfaces (BCI), facilitating direct interaction between the human brain and external devices \cite{kawala2021summary}, and enjoys acceptance in the consumer market, especially for gaming and health-related applications.

Identity verification, commonly known as \textbf{authentication}, involves a one-to-one comparison to confirm if an individual presenting a previously registered biometric trait is indeed the same person. In essence, it addresses the question: "\textit{Are you who you claim to be?}" This process is crucial to differentiate legitimate users from imposters. Biometric authentication comprises two primary phases: enrollment and verification~\cite{ISO2382-37}. During enrollment, users' unique biometric characteristics are captured and stored as templates or used to train and store a classification model for that user. Each template/user model is linked to a unique user identifier, such as a username. In the verification phase, the system compares the current biometric sample with the stored template or runs it through the classification model, corresponding to the claimed identity.

\begin{figure} 
    \centering
    \includegraphics[width=1\linewidth]{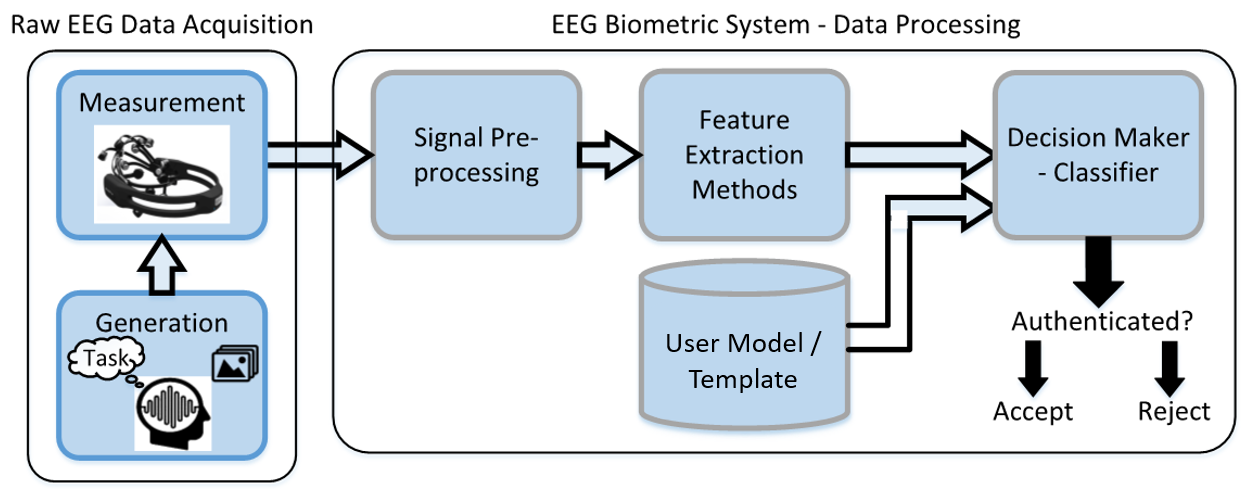}
    \caption{Overview of elements in a brainwave-based biometric authentication process ~\cite{arias2023performance}.}
    \label{fig:EEG_Au}
\end{figure}

The \textbf{brainwave-based authentication process}, depicted in Figure \ref{fig:EEG_Au}, unfolds through several key stages:

\begin{itemize}
    \item \textbf{Raw EEG Data Acquisition:} This initial step captures the brain's electrical activity. An effective way to elicit signals useful for authentication is measuring brain reactions to controlled stimuli, such as audio or images. These time-locked reactions, called Event-Related Potentials (ERPs), have shown unique features to individuate people \cite{arias2021inexpensive} and are distinguished by their high Signal-to-Noise Ratio (SNR)~\cite{armstrong2015brainprint,zhang2021review}. Moreover, the ability to modify the stimuli associated with ERPs enhances the system's revocability. This feature is invaluable in mitigating risks associated with potential data compromise~\cite{lin2018brain}. NeuroIDBench focuses in ERP-based datasets, given their superior benefits with regard to alternative brainwave elicitation paradigms.

    \item \textbf{Signal Pre-processing:} The acquired raw EEG data is then refined through pre-processing. This may include filtering out noise, rejecting artifacts, and downsampling the signal, aiming to isolate the purest form of the EEG data for authentication purposes~\cite{gui2019survey}. NeuroIDBench embeds a pre-processing pipeline based on best-practices for handling ERPs as used in the community \cite{gui2019survey}.

    \item \textbf{Feature Extraction:} The pre-processed EEG signals are further analyzed to extract key features. These features can be directly obtained from the ERP EEG signal, such as power spectral densities, or generated as representation vectors learned by a neural network. NeuroIDBench implements both kind of approaches.
    
    \item \textbf{User Model/Template Database:} This Database stores the user model learned from the collected features in order to perform classification, or the template of features representing the user. 
    
    \item \textbf{Decision Maker:} In the authentication phase, the system evaluates the incoming EEG sample, 
     against the stored model/template, applying similarity thresholds to decide if it belongs to a legitimate user or to an impostor. NeuroIDBench incorporates the most common techniques for machine learning model-based approaches (SVM, KNN, LDA, NB, LR, RF), as well as the main approach for template-based authentication using deep neural networks (Twin Neural Networks). 
\end{itemize}

\subsection{Adversary Model}
In addressing the security of brainwave-based authentication systems, we consider the zero-effort attacker as the primary threat model. This approach aligns with the standard threat model used in most brainwave authentication studies. A zero-effort attack is characterized by an adversary who attempts to gain unauthorized access using their own biometric data, without employing any advanced techniques to mimic or falsify another individual's biometric trait~\cite{mansfield2002best}. In this context, the key challenge in protecting against zero-effort attacks lies in the biometric system's ability to accurately distinguish between the biometric data of an actual user and that of the intruder. The system must be adept at minimizing false matches, where the adversary's biometric data is erroneously accepted as that of a legitimate user.

In our consideration of zero-effort attacks, we distinguish between two types of attackers: \textit{known} and \textit{unknown}. A known attacker refers to one whose biometric data has previously been exposed to the system, typically during the training phase. This exposure can influence the system's learning and adaptation, potentially affecting its ability to accurately identify or reject these attackers later. On the other hand, an unknown attacker is one whose biometric data has never been introduced to the system. This lack of prior exposure means that the system has no learning or adaptation based on this attacker's data, making it a more realistic and challenging scenario. We expect that unknown attackers represent a higher risk to the system, as they test the system's ability to authenticate users based purely on learned patterns without prior knowledge of the attacker. While the unknown attacker scenario is realistic and should be considered in evaluations, the known attacker scenario has been used in some papers. Therefore, we provide results for both to demonstrate how unrealistic attacker scenarios can influence outcomes and to encourage researchers to use the unknown attacker scenario.

\begin{figure*}[!ht]
    \centering
    \includegraphics[width=1\textwidth]{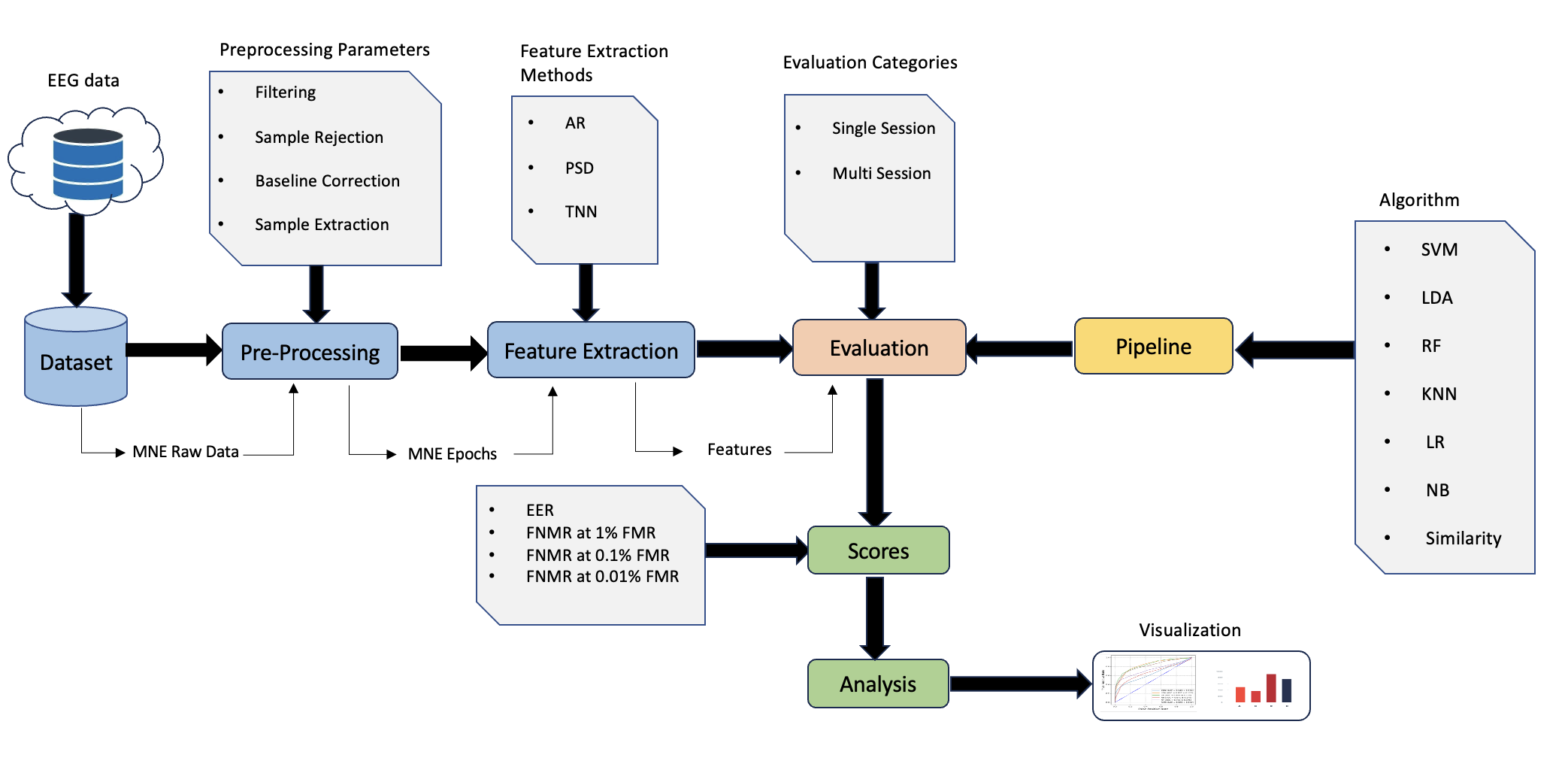}
    \caption{Schematic depiction outlining the architectural structure of the NeuroIDBench codebase, providing a visual overview of its underlying framework \cite{jayaram2018moabb}}
    \label{fig:overview}
\end{figure*}

\subsection{Related Work}

\begin{table}[!h]
\centering
\caption{\small Comparison of Related Work on Brainwave-Based Authentication, detailing Dataset Identity Population (D.P.), Dataset Availability (D.A.), and Code Availability (C.A). (For papers that had more than one independent dataset. Population of them split with comma (",") )}
\begin{tabular}{lccccccc}
\hline
 \textbf{Publication} & \textbf{D.P.}  & \textbf{D.A.} & \textbf{C.A.} \\ \hline
Arias et al.(2023) \cite{arias2023performance} & 52,40  &  \checkmark & $\times$ \\
Fallahi et al.(2023) \cite{fallahi2023brainnet} & 41,40  &  \checkmark & \checkmark\\
Hernandez  et al.(2022) \cite{hernandez2022eeg}& 39 &  $\times$ &  $\times$   \\
Bidgoly  et al.(2022) \cite{bidgoly2022towards}& 109 &  \checkmark &  $\times$   \\
Maiorana(2021) \cite{maiorana2021learning}& 45 & $\times$  & $\times$ \\
Sooriyaarachchi  et al.(2020) \cite{sooriyaarachchi2020musicid}& 20 &  $\times$ &  $\times$   \\
Gupta  et al.(2020) \cite{gupta2020blink}& 20 &  \checkmark &  \checkmark   \\
Nakanishi et al.(2019) \cite{nakanishi2019biometric}& 10 &  $\times$ &  $\times$  \\
Lin et al.(2018) \cite{lin2018brain}& 179 & $\times$  &  $\times$  \\
Schons et al.(2018) \cite{schons2018convolutional}& 109 &  \checkmark &  $\times$  \\
Das et al.(2016) \cite{das2016eeg} & 50 &  $\times$ &  $\times$  \\
\hline

\end{tabular}
\label{tab:related}
\end{table}

Presently, two primary methods for brainwave authentication are studied in the literature. The first method utilizes shallow classifiers built on time series-based feature extraction techniques, such as Power Spectral Density (PSD) or Autoregressive (AR) models \cite{arias2023performance,arias2021inexpensive,nakanishi2019biometric,das2016eeg}. The second approach involves training specialized deep learning models designed explicitly for brainwave data, aiming to extract features directly associated with user identity~\cite{fallahi2023brainnet,maiorana2021learning,schons2018convolutional,bidgoly2022towards}.
However, traditional feature extraction methods are not sufficiently effective, as they were not originally designed for authentication purposes. Meanwhile, deep learning approaches require a substantial amount of data to develop an adequate feature extractor, which is currently scarce in the field of brainwave authentication. Therefore, it is crucial to determine the most appropriate hyperparameters based on other datasets in the field, which can aid in mitigating these challenges.

Table \ref{tab:related} presents an analysis of 11 papers in the realm of brainwave authentication. Among these, only two papers have made their source code available~\cite{fallahi2023brainnet,gupta2020blink}, and merely five have utilized public datasets \cite{fallahi2023brainnet,gupta2020blink,arias2023performance,bidgoly2022towards,schons2018convolutional}. Notably, four of these papers employed two common public datasets \cite{erpcore,schalk2004bci2000}, reflecting the limited data availability, which consequently diminishes the generalizability of the results. On average, these studies included approximately 70 subjects for their evaluation purposes. The subcommunity in brainwave-based authentication clearly require more transparency in the research practice as well as more accessibility in the datasets and source code. To be sure, the brainwave-authentication subcommunity are, in this regard, not alone. For example, Olszewski et al.~\cite{olszewski2023get} find that the "large and long-lived" subcommunity of machine learning security fail to provide data (30\%) or source code (40\%) even in the papers published at the top-tier security conferences. The authors found out that a full 80\% of artifacts are incapable of reproducing the reported results. It is therefore unsurprising that similar problems should plague the new and emerging subcommunity in brainwave-based authentication. Nevertheless, these issues are intensified in brainwave authentication due to the relatively small size of each dataset, heightening the risk of overfitting.

Meanwhile, researchers in the field of Brain-Computer Interface (BCI) encounter similar challenges. Hence, Jayaram and Barachant~\cite{moabb} proposed a benchmark framework\footnote{https://github.com/NeuroTechX/moabb} for investigating common BCI algorithms across 22 public EEG datasets with over 250 participants. This framework aims to address issues of reproducibility, lack of open-source resources, and result compatibility. However, their study did not cover authentication algorithms. Therefore, it is vital for brainwave authentication research to adopt an open-source approach and to extensively employ existing public datasets. It is crucial to take such steps to enhance the robustness of research in this field and to promote a deeper, more comprehensive exploration of brainwave authentication techniques.

\begin{table*}[!ht]
\captionsetup{font=small} 
\centering
\caption{\small Publicly available ERP datasets based on N400 (Semantic Priming) paradigm and P300 (oddball) paradigm}
\begin{tabular}{cccccccccc}
\hline
\textbf{Paradigm} &\textbf{Dataset} & \textbf{Year} & \textbf{Subjects} & \textbf{EEG Device} & \textbf{Channels} & \textbf{S.Rate}& \textbf{Sessions} \\\hline

N400&Pijnacker \textit{et al.} \cite{pijnacker2017semantic} & 2017 & 45 & actiCap & 32 & 500 Hz & 1  \\
N400 &Draschkow \textit{et al.} \cite{draschkow2018no} & 2018 & 40 & BrainAmp, actiChamp & 64 & 1000 Hz & 1 \\
N400 &Marzecová \textit{et al.} \cite{marzecova2018attentional} & 2018 & 18 & BrainAmp & 59 & 500 Hz & 1 \\
\textbf{N400} &\textbf{Mantegna \textit{et al.} \cite{mantegna2019distinguishing}} & \textbf{2019} & \textbf{31} & \textbf{BrainAmp, EasyCap} & \textbf{65} & \textbf{500 Hz} & \textbf{1} \\

\textbf{N400} &\textbf{ERPCORE: N400 \cite{erpcore}} & \textbf{2021} & \textbf{40} & \textbf{Biosemi} & \textbf{30} & \textbf{1024 Hz} & \textbf{1} \\
N400 &Hodapp and Rabovky \cite{hodapp2021n400} & 2021 & 33 & BrainAmp & 64 & 1000 Hz & 1  \\
N400 &Rabs \textit{et al.} \cite{rabs2022situational} & 2022 & 38 & BrainVision & 26 & 500 Hz & 1  \\
N400 &Schoknecht \textit{et al.} \cite{schoknecht2022interaction} & 2022 & 38 & ActiCap, ActiChamp & 58 & 500 Hz & 1  \\
N400 &Toffolo \textit{et al.} \cite{toffolo2022evoking} & 2022 & 24 & Biosemi & 128 & 512 Hz & 1 \\
N400 &Lindborg \textit{et al.} \cite{lindborg2023semantic} & 2022 & 40 & BrainVision & 64 & 2046 Hz & 1 \\
\textbf{N400} & \textbf{COGBCI: Flanker \cite{cogbci}} & \textbf{2023} & \textbf{29} & \textbf{ActiCap, actiChamp} & \textbf{64} & \textbf{512 Hz} & \textbf{3} \\
N400 &Stone \textit{et al.} \cite{stone2023understanding} & 2023 & 64 & TMSi Refa & 32 & 512 Hz & 1 \\ \hline

P300 & BrainInvaders12 \cite{van2019building} & 2012 & 25 & NeXus-32 & 16 & 128 Hz & 1  \\
P300 & BrainInvaders13a \cite{vaineau2019brain} & 2013 & 24 & g.GAMMAcap & 16 & 512 Hz & 1 \\
P300 & BrainInvaders14a \cite{bi2014a} & 2014 & 64 & g.Sahara & 16 & 512 Hz & 1  \\
P300 & BrainInvaders14b \cite{bi2014b} & 2014 & 37 & g.GAMMAcap & 32 & 512 Hz & 1 \\
P300 & Gao \textit{et al.} \cite{gao2014novel} & 2014 & 30 & Neuroscan & 12 & 500 Hz & 1  \\
\textbf{P300} & \textbf{BrainInvaders15a \cite{brainInvaders15a}} & \textbf{2015} & \textbf{50} & \textbf{g.GAMMAcap} & \textbf{32} & \textbf{512 Hz} & \textbf{1}  \\
P300 & BrainInvaders15b \cite{bi2015b} & 2015 & 44 & g.GAMMAcap & 32 & 512 Hz & 1  \\
P300 & Mouček \textit{et al.} \cite{mouvcek2017event} & 2017 & 250 & BrainVision & 3 & n.a. & 1  \\
\textbf{P300} & \textbf{Hubner \textit{et al.} \cite{hubner2017learning}} & \textbf{2017} & \textbf{13} & \textbf{BrainAmp DC} & \textbf{31} & \textbf{1000 Hz} & \textbf{1}  \\
\textbf{P300} & \textbf{Sosulski and Tangermann \cite{sosulski2019spatial}} & \textbf{2019} & \textbf{13} & \textbf{BrainAmp, EasyCap} & \textbf{31} & \textbf{1000 Hz} & \textbf{1} \\
\textbf{P300} & \textbf{Lee \textit{et al.} \cite{lee2019eeg}} & \textbf{2019} & \textbf{54} & \textbf{BrainAmp} & \textbf{62} & \textbf{1000 Hz} & \textbf{2}  \\
P300 & Simões \textit{et al.} \cite{simoes2020bciaut} & 2020 & 15 & g.tec & 8 & 250 Hz & 7  \\
P300 & Goncharenko \textit{et al.} \cite{goncharenko2020raccoons} & 2020 & 60 & NVX-52 & 8 & 500 Hz & 1  \\
P300 & Chatroudi \textit{et al.} \cite{houshmand2021effect} & 2021 & 24 & g.tec & 64 & 1200 Hz & 1  \\
P300 & Cattan \textit{et al.} \cite{cunningham2021underestimation} & 2021 & 21 & g.USBamp, g.tec & 16 & 512 Hz & 1  \\
\textbf{P300} & \textbf{ERPCORE: P300 \cite{erpcore}} & \textbf{2021} & \textbf{40} & \textbf{Biosemi} & \textbf{30} & \textbf{1024 Hz} & \textbf{1} \\
\textbf{P300} & \textbf{Won \textit{et al.} \cite{won2022eeg}} & \textbf{2022} & \textbf{55} & \textbf{Biosemi} & \textbf{32} & \textbf{512 Hz} & \textbf{1} \\ \hline
\end{tabular}
\label{tab:erp_datasets}
\end{table*}

\section{NeuroIDBench}
\label{sec:method}
As illustrated in Figure \ref{fig:overview}, NeuroIDBench is organized into five integral components: dataset, preprocessing, feature extraction, evaluation, and analysis.  Our benchmarking workflow is influenced by MOABB (Mother of all BCI benchmarks) \cite{jayaram2018moabb} work. 
In line with our research aims, we have tailored and refined their methodology to create a customized benchmarking suite specifically designed for evaluating and analyzing brainwave authentication systems. In the following, we detail NeuroIDBench's components, justifying the design choices.

    \subsection{Dataset Loading}
    In response to the need for flexible dataset management in EEG research, we have developed an interface aimed at streamlining the handling of diverse datasets with various formats, thereby reducing the burden on researchers. In the past, this field has been constrained by the laborious task of manually importing and formatting each dataset individually, often limiting researchers to using only one or two datasets in their analyses due to these logistical challenges. To overcome this, our solution is a straightforward Python component that automatically downloads and processes nine distinct EEG datasets using the MNE Python package\footnote{https://mne.tools/stable/index.html}, a recognized and open-source brainwave library. This tool not only simplifies the integration of these existing datasets but also allows for the easy adoption of new datasets into the benchmark by following established examples and specifying the relevant event IDs. The expected outcome of this innovation is a significant enhancement in the use of diverse datasets in research, promoting the avoidance of overfitting to specific datasets and fostering more generalizable and robust scientific findings. This development represents a leap forward in EEG-based research, offering researchers a more efficient and comprehensive approach to dataset management and analysis.

    \label{sec:dataset}
    In choosing our dataset, we focused on Event-Related Potential (ERP) paradigms, particularly the P300 and N400, which are the most well-known and easier to elicit.
     The P300 reaction is triggered with the `oddball' paradigm, i.e., including a rare stimulus in a sequence \cite{picton1992p300}, while the N400 is linked to semantic processing \cite{debruille1996n400}. Our selection methodology involved a thorough review of publicly accessible datasets utilizing these ERP paradigms. We carefully examined and assessed over 29 datasets (see Table \ref{tab:erp_datasets}). Ultimately, we chose 9 datasets for further investigation. This selection included two multi-session datasets to examine performance over multiple sessions \cite{cogbci,lee2019eeg}, five datasets with over 30 subjects to evaluate performance in datasets with a relatively large number of samples \cite{mantegna2019distinguishing,erpcore,brainInvaders15a,won2022eeg}, and two datasets with just 13 subjects to explore performance in smaller datasets \cite{hubner2017learning,sosulski2019spatial}. 
     
       Our prioritization focused on datasets providing raw (unprocessed) data with ample event information, facilitating the extraction of relevant P300 or N400 samples. This emphasis led to the implementation of a condition favoring datasets with raw data, allowing us to uniformly apply standardized pre-processing, feature extraction, and authentication steps across all selected datasets. This uniform procedure is crucial for evaluating their performance under comparable experimental conditions, an unattainable task without access to unprocessed raw data. Consequently, datasets offering solely pre-processed data were excluded from our study. Additionally, the necessity for comprehensive event information was paramount; without it, the assurance of working with the correct events, intricately associated with the elicitation of P300 and N400, would have been compromised.

    \subsection{Pre-processing}
    In this module, we conduct pre-processing on the standardized MNE Raw data. Various methods exist for cleansing artifacts; however, the procedures must remain consistent to ensure the validity of comparisons between algorithms or datasets \cite{jayaram2018moabb}. We adhere to established best practices commonly employed in pre-processing methodologies within brainwave authentication studies \cite{gui2019survey}. To cleanse EEG artifacts, we begin by implementing bandpass filtering in the range of 1 to 50 Hz.  
    Subsequently, we extract samples from the raw signals, temporally aligning the data to a range spanning from -200 to 800 ms relative to the onset of the stimulus. Afterward, Following the baseline correction, where we subtract the mean of a baseline period (-200 to 0 ms) from the entire sample data points to mitigate drift effects like DC offsets \cite{fallahi2023brainnet}. Finally, we employ a peak-to-peak rejection method to eliminate large artifacts attributed to eye or muscle movements from contaminating the EEG data.  
    
    While adhering to established pre-processing standards prevalent in brainwave authentication studies, we recognize the necessity for researchers to tailor pre-processing approaches to suit their specific EEG data requirements. To address this need for flexibility, we designed this pre-processing interface with customization options. These functionalities empower researchers to adjust parameters, including the selection of sample interval, threshold settings for sample rejection, and the application of baseline correction, making it a more adaptable tool.

   \subsection{Feature Extraction}
   This module is designed to extract features from pre-processed EEG data. It specifically focuses on identifying characteristics in both the temporal and frequency domains. More precisely, it uses Autoregressive (AR) coefficients to represent time-domain properties and calculates Power Spectral Density (PSD) to capture frequency domain features.  These calculated features, namely AR coefficients and PSD, are subsequently harnessed as inputs for facilitating authentication through shallow classifiers. In contrast, for similarity-based authentication, the module employs the Twin Neural Network (TNN) to generate feature embeddings. These embeddings, produced by the TNN, play a pivotal role as inputs for similarity-based authentication approaches. 
    
    To compute AR coefficients, our system fits AR models to the pre-processed samples, employing a one-second duration of time series data \cite{arias2023performance}. The estimation of these AR coefficients utilized the Yule-Walker method \cite{pardey1996review}. Determining the ideal order for AR modeling poses a challenge, as higher orders escalate computational demands, while exceedingly low orders inadequately represent the signal \cite{zhang2017classification}. To address this, our system allows users the flexibility to define their preferred AR order, enabling a tailored approach aligned with their specific requirements.

    The PSD of each sample is computed across different frequency bands, namely low (1-10 Hz), $\alpha$ (10-13 Hz), $\beta$ (13-30 Hz), and $\gamma$ (30-50 Hz), utilizing the Welchs periodogram algorithm \cite{arias2023performance}. The calculation involved first determining the PSD for each frequency point in the 1-second sample. For our PSD computation, we used four time windows that were all the same size. These windows were applied to a 1-second ERP sample, with a 50$\%$ overlap between each window. The inclusion of the time window factor was crucial to segregate the authentic frequency modulation of the EEG induced by attention from any artifacts that the attentional modulation of ERPs might have induced \cite{comez1998frequency}. Subsequently, we computed the average PSD within the specified frequency ranges, enabling us to determine the average power spectrum of the low, $\alpha$, $\beta$, and $\gamma$ frequency bands.

\begin{figure*}[!ht]
    \centering
    \includegraphics[width=\textwidth]{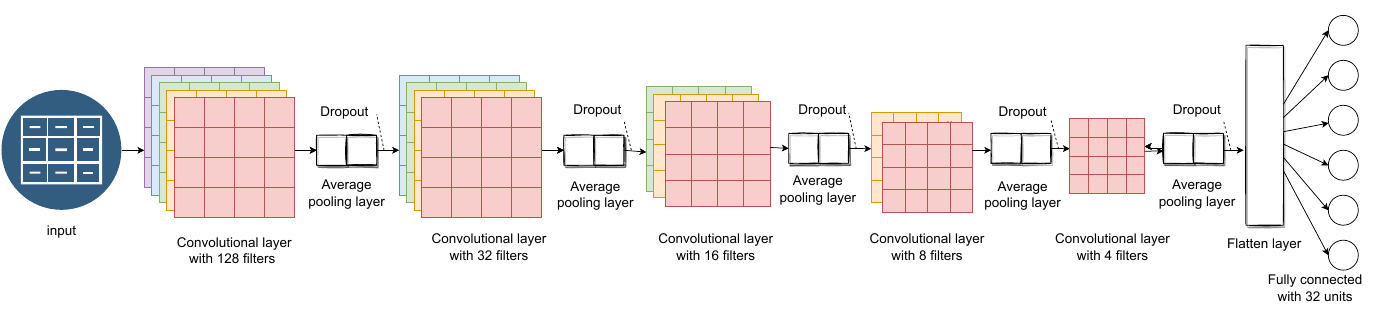}
    \caption{The CNN's network architecture within the Twin neural Network is designed to produce a condensed 32-bit embedding from brain data samples, serving as an efficient method to compute latent representations of inputs \cite{fallahi2023brainnet}.} 
    \label{fig:Siamese Neural Network Architecture}
\end{figure*}

    Furthermore, NeuroIDBench implements a deep learning based approach using a TNN with triplet loss function to transform time-series EEG data into compact brain embeddings. Building on the work of Fallahi \textit{et al.} \cite{fallahi2023brainnet}, our TNN structure integrates three Convolutional Neural Network (CNN) branches, each featuring five convolutional layers, as illustrated in Figure \ref{fig:Siamese Neural Network Architecture}. These branches received 1-second samples as input, structured in a two-dimensional array where rows correspond to channel indices, and columns represent discrete-time measurements. After each convolutional layer, we applied an average pooling layer to reduce the input vectors' dimensionality while preserving the unique characteristics of each brainwave.

    However, acknowledging the diverse nature of EEG datasets and the variability in research requirements, we have extended the flexibility of our framework; To accommodate varying preferences and dataset characteristics, users can conveniently adjust critical training parameters such as batch size, the number of epochs for training, and the learning rate within the tool. Also, the framework allows users to integrate their neural network architectures. This flexibility accommodates diverse methodologies for brainwave authentication studies, empowering researchers to assess and compare various deep-learning approaches across a wide range of EEG datasets within our tool.

    \subsection{Evaluation}
    While most EEG studies assess brainwaves in a single-session scenario \cite{yap2023evaluation,fallahi2023brainnet,arias2023performance,bidgoly2022towards,schons2018convolutional,das2016eeg}, it is crucial to examine EEG robustness over time. A realistic and often overlooked approach is the multi-session scenario. To highlight this gap, we provide both single-session and multi-session evaluation scenarios in our benchmark tools as follows
    
    \textbf{Single-Session Evaluation:} Under the single-session evaluation, the training and testing of the features are done utilizing the recorded data from a single session. To avoid overfitting and increase the reliability of our authentication system, cross-validation schemes have been employed in both known and unknown attacker scenarios. 
    
    For the known attacker scenario, we implemented a Stratified cross-validation approach with k=4 to divide the single-session data into training and testing sets. This choice ensures the representation of features from both classes in both training and testing data across each fold. We take the step of eliminating users with fewer than four samples from the datasets to guarantee sufficient samples for both training and testing \cite{arias2023performance}. Conversely, the unknown attacker approach adopts the GroupKFold cross-validation strategy with k=4, where grouping is based on SubjectID, resulting in non-overlapping training and testing sets for users in each cross-validation round. Evaluation metric results are aggregated across all folds, providing averaged and comprehensive reports. In datasets like COGBCI and Lee2019, which had multiple sessions, we evaluate each session separately, subsequently averaging the results from all the sessions as single-session evaluation results. Multi-session evaluation, described later in this section.
        
    Additionally, we actively apply z-scaler normalization to both training and testing data to address potential issues arising from variations in feature scales, ensuring equitable contribution from all features during model training \cite{hastie2009elements}. This approach aligns with established best practices in machine learning \cite{pedregosa2011scikit}, reinforcing the reliability and generalization capability of our EEG-based authentication system.

    \textbf{Multi-Session Evaluation:} The multi-session evaluation methodology was applied to two EEG datasets, namely COGBCI~\cite{cogbci} and Lee2019~\cite{lee2019eeg}. COGBCI consists of three sessions separated by day intervals, while Lee2019 includes two sessions separated by week intervals. A cross-validation strategy was employed to address overfitting. This involved dividing the attacker subjects for the shallow classifiers and dividing the subject of latent space learning and evaluation.
    
    In this setup, if we denote the EEG samples from the ith session as \(X_i\), enrollment is being conducted using \(X_i\), and authentication is performed using \(X_{j}\), where \(j>i\). For instance, if a dataset had three sessions (i.e., \(i = 1, 2, 3\)), authentication is performed using \(X_2\) and \(X_3\) after enrolling with \(X_1\) and similarly using \(X_3\) as authentication samples after enrolling with \(X_2\). This approach ensures that the model is trained on earlier sessions' data and tested on subsequent sessions, mimicking the real-world scenario where user enrollment precedes authentication. Notably, this logical sequence aligns with common practices in practical biometric systems. The reported results are the average of the results from all the testing sessions. 

    Similar to the single-session evaluation approach, we implemented data normalization on both training and testing datasets. Specifically, we applied 'fit and transform' to the training data, while only 'transform' was used for the testing data. Evaluation metrics were computed for each cross-validation fold and then averaged across sessions for reporting.

    \subsection{Analysis}
    Proper metrics are essential for comparing different methods from the viewpoint of uniqueness. We have chosen to report EER (Equal Error Rate) and FNMR (False Non-Match Rate) at various FMR (False Match Rate) levels, as recommended by the biometric authentication community~\cite{eberz2017evaluating,sundararajan2019survey} and international standards \cite{mansfield2006information,grassi2020digital}. The FMR indicates the likelihood of an unauthorized user being mistakenly authenticated by the system, reflecting the system's vulnerability to a zero-effort attack. The goal is to achieve a low FMR while maintaining an acceptable FNMR. A high FNMR can result in legitimate users being repeatedly denied access, negatively impacting the device's usability. Therefore, it is crucial to balance FMR and FNMR to ensure robust security without compromising user experience. The EER represents the point, threshold, where the FMR and FNMR are equal.

Furthermore, this benchmarking framework is developed with a primary focus on ensuring ease of use. Our goal is to enable users to efficiently utilize the framework, even without a thorough grasp of the complex technical aspects of the Python programming language. To accomplish this, we have created a user-friendly benchmarking script that proficiently examines a configuration file written in a straightforward YAML interface. This configuration file serves as a control panel, enabling users to specify a wide range of parameters and settings. This system effortlessly automates the complex processes of data extraction, pre-processing, feature extraction, and classification, as illustrated in Figure \ref{fig:overview}. This streamlined approach eliminates the need for users to delve into intricate programming complexities. In \ref{Appendix Installation}, the procedural steps for tool installation are detailed, accompanied by illustrative examples of configuration files for the automated creation of brainwave authentication pipelines.

\section{Benchmarking Results and Discussion}
\begin{figure*}
    \centering
    \includegraphics[width=1.0\linewidth]{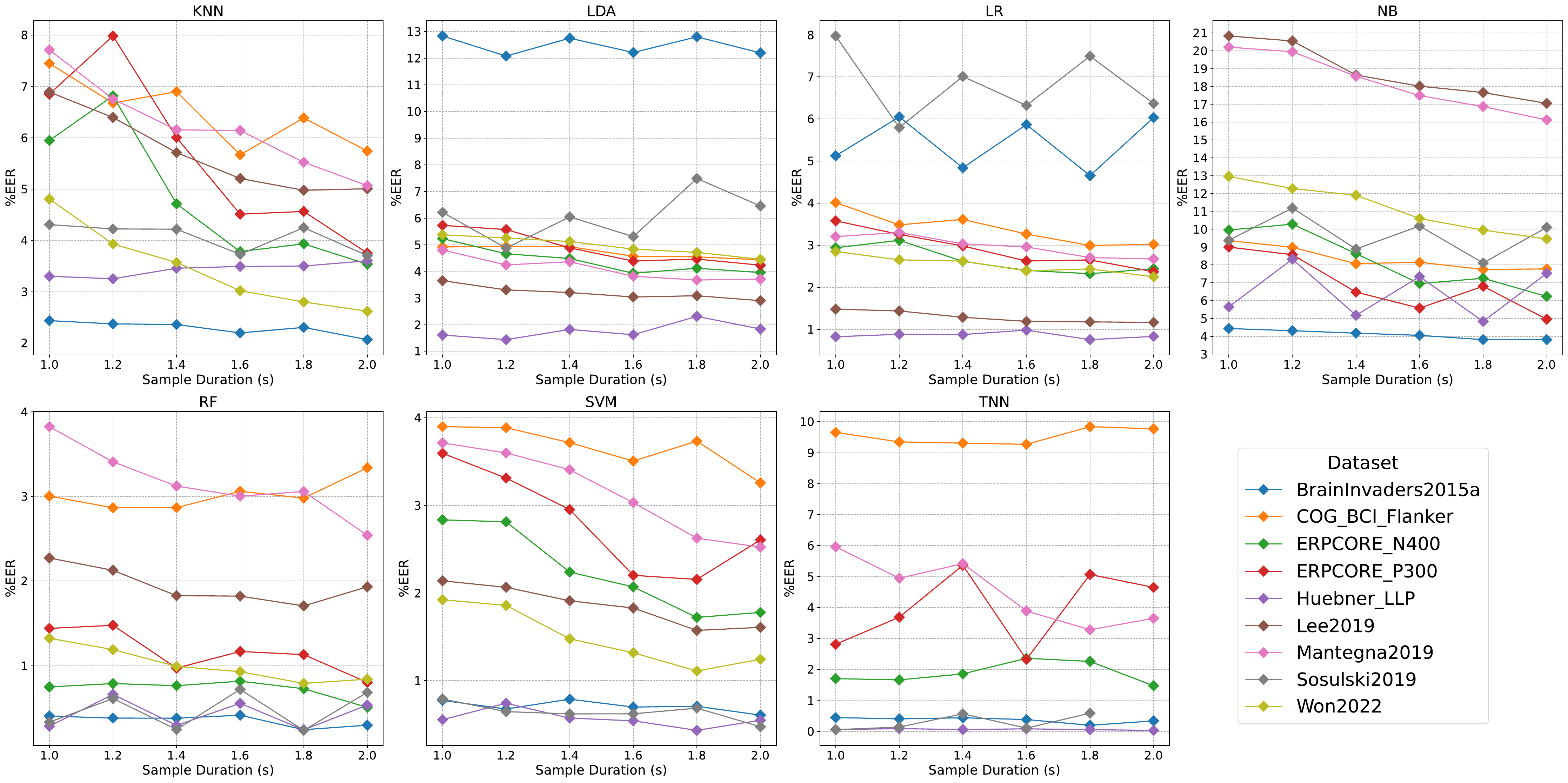}
    \caption{Impact of varying sample duration, ranging from 1 to 2 seconds, on the efficacy of the seven authentication algorithms. These algorithms are applied across nine datasets, with the evaluation metric employed being the EER. The evaluation was conducted under the unknown attacker scenario.} 
    \label{fig:Epochs_Duration}
\end{figure*}

We began with preprocessing and feature extraction to identify the best parameters for subsequent experiments. Subsequently, we compared known and unknown attacker models to ascertain the increased difficulty associated with unknown problems. This was followed by an investigation into single and multi-session performances with unknown attacker scenarios. Additionally, we examined the effectiveness of two primary approaches: shallow classifiers and similarity approaches.

\subsection{Preprocessing}
\label{sec:pre}
This study examines two critical aspects of electroencephalography (EEG) preprocessing: sample duration and sample rejection. Typically, researchers utilize a one-second sample, spanning from 0.2 seconds before to 0.8 seconds after a stimulus. We explore various sample durations, specifically [1.0, 1.2, 1.4, 1.6, 1.8, 2.0] seconds. It is hypothesized that longer durations may enhance results due to the inclusion of more information. However, considering the primary objective of event-related potentials (ERP) is to introduce stimuli for signal-to-noise ratio (SNR) enhancement, extended samples might inadvertently incorporate noise data. Another preprocessing step investigated is sample rejection, based on peak-to-peak (PTP) signal amplitude. We establish a minimum acceptable threshold for signal amplitude, rejecting samples that fall below this threshold. Continuing from the previous section, the rationale behind PTP sample rejection is the assumption that samples with high PTP values are likely to be noisy and their removal might enhance the performance of brainwave-based authentication systems. Our investigation focuses on determining whether this preprocessing step can indeed improve the efficacy of brainwave authentication and, if so, identifying the optimal threshold level for PTP sample rejection.

Figure \ref{fig:Epochs_Duration} presents the EER results for the sample duration experiment. The results indicate that longer samples may enhance outcomes in shallow classifiers, but no significant improvement was observed in the case of Twin neural networks. It is important to note that for longer samples (e.g., 2 seconds), it was necessary to downsample the sample to mitigate the increased data dimensionality, a step taken due to hardware limitations as suggested in source code of BrainNet \cite{fallahi2023brainnet} paper as well.

\begin{figure*}
    \centering
    \includegraphics[width=1.0\linewidth]{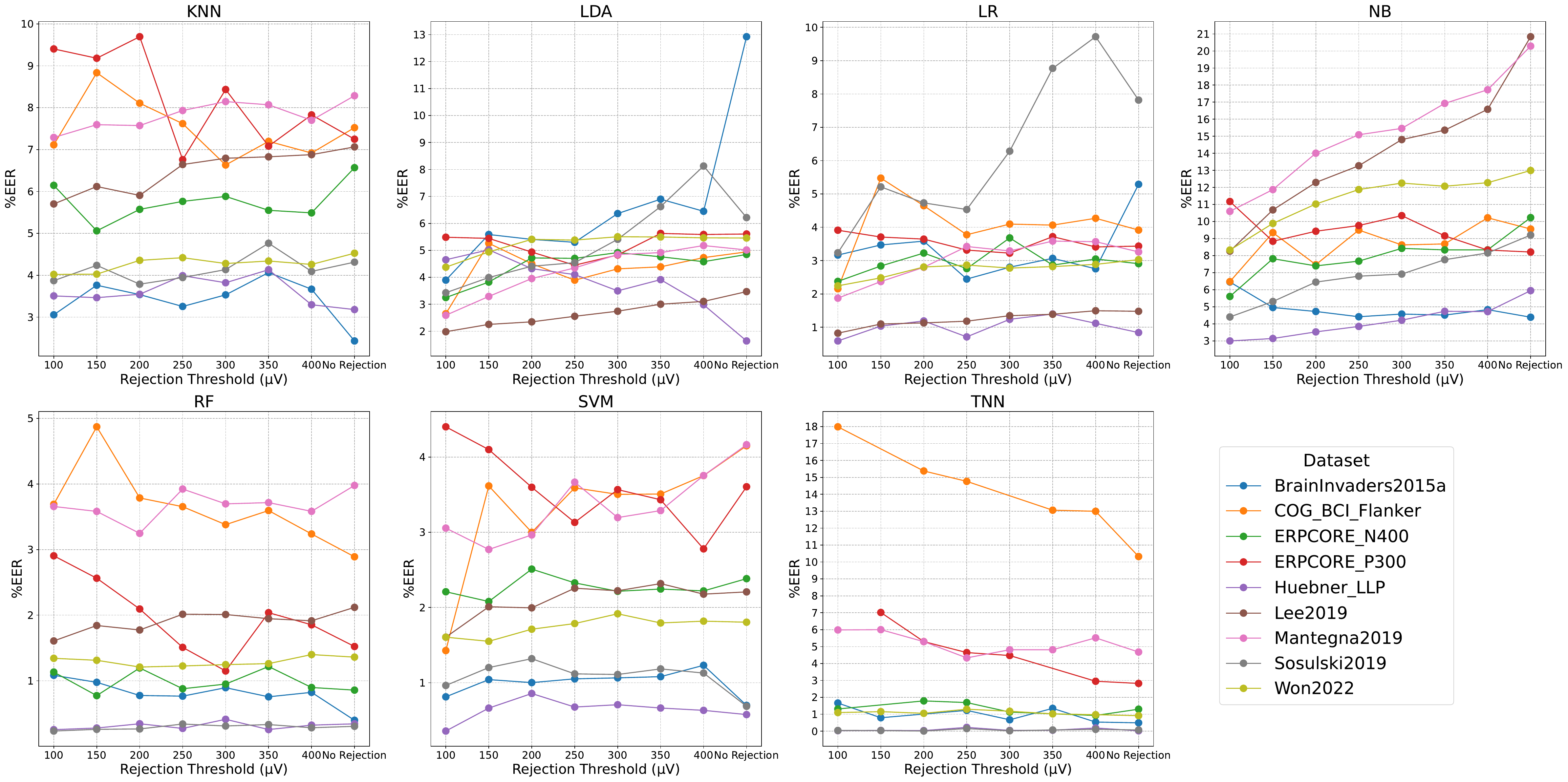}
    \caption{Influence of varying sample rejection thresholds, spanning from 100 to 400 microvolts, as well as scenarios with no rejection, on the performance outcomes of seven authentication algorithms. The investigation is conducted across nine distinct datasets, employing the EER as the metric for comprehensive performance assessment, and performance assessment is done under an unknown attacker scenario.}
    \label{fig:Epochs_rejection}
\end{figure*}

Figure \ref{fig:Epochs_rejection} displays EER across various PTP rejection thresholds. Results show that for classifiers like LR, LDA, and NB, a lower threshold results in a reduced EER. In contrast, RF and SVM displayed no specific trend, and their performance without sample rejection already surpassed other shallow classifiers. For the Twin Neural Network, sample rejection had either no effect or a negative impact. Overall, the study found no significant improvement attributable to PTP rejection.

\begin{figure*}
    \raggedright
    \includegraphics[width=0.89\linewidth, left]{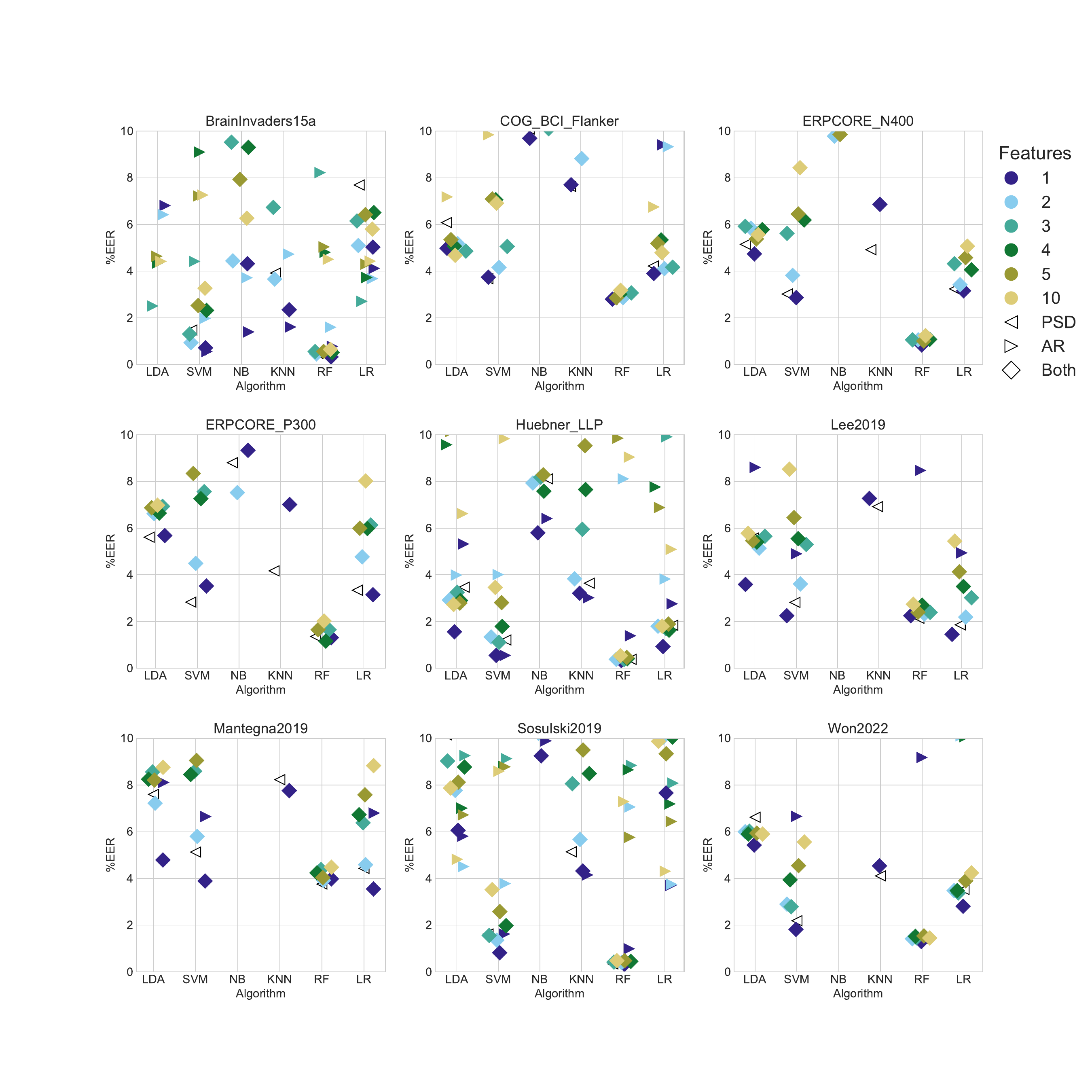}
    \caption{Comparative performance evaluation of nine datasets under unknown attacker scenario: Employing shallow classifiers and time-frequency domain features, evaluated by EER.  
    }
    \label{Feature_Extraction}
\end{figure*}
\subsection{Feature Extraction}
\label{sec:feature}
We explored Power Spectral Density (PSD) and Autoregressive (AR) models of different orders as feature extraction methods typically employed in shallow classifiers. Figure \ref{Feature_Extraction} illustrates the EER for various configurations of feature extraction across different classifiers. The findings suggest that the combination of PSD with AR of order 1 yields superior performance compared to other combinations. Following this, PSD features alone demonstrate promising results, whereas AR on its own fails to show stable and robust outcomes. Interestingly, in the BrainInvaders15a dataset, the AR of order 1 outperformed most classifiers. In the ERPCORE\_P300 dataset, PSD was the predominant feature leading to superior performance across most classifiers. Additionally, in the COG\_BCI dataset, a higher AR order demonstrated better performance compared to lower AR orders. These results substantiate our assertion that analyzing a single dataset can yield results that are not generalizable to other datasets. However, it is noteworthy that across all these datasets, the combination of PSD with AR order 1 consistently appears among the top-performing features. Consequently, we will utilize these features in the subsequent sections of this paper as the default feature set for shallow classifiers.
\begin{figure*}
    \centering
    \includegraphics[width=1\linewidth, height=0.3\linewidth]{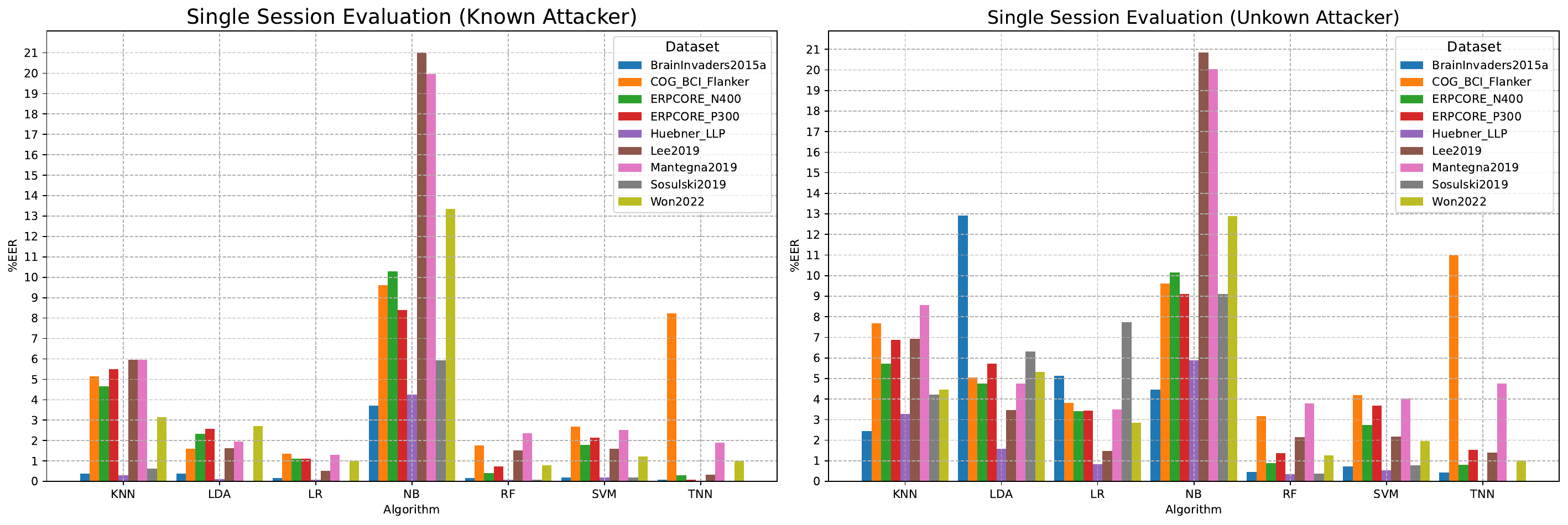}
    \caption{Examining the effects of known and unknown attacker models on EER in brainwave authentication across various classifiers and datasets}
    \label{fig:Session_Evaluation}
\end{figure*}
\subsection{Known and UnKnown Attacker Approach}
\label{sec:seen}
The known attacker model represents an unrealistic scenario; however, given its existence in some previous studies, we investigated the performance gap between known and unknown attacker scenarios. Figure \ref{fig:Session_Evaluation} displays the EER for these two scenarios, based on a sample duration of 1 second to ensure comparability with previous work. This analysis was conducted without sample rejection and utilized PSD combined with AR order 1 as the feature extraction method for shallow classifiers.

The results reveal that the mean EER across datasets degraded by 58.44\% for KNN, 275.60\% for LDA, 383.91\% for LR, 5.83\% for NB, 75.94\% for RF, 66.61\% for SVM, and 75.88\% for the Twin neural network approach. Notably, in some cases, the EER increased several-fold. These findings suggest that results derived from the known attacker model can be misleading, emphasizing the need for researchers to exercise caution and potentially avoid using this model in their analyses.

\subsection{Single Session vs. Multi Session Authentication}
\label{sec:session}
For the practical application of brainwave authentication, it is essential to develop a model capable of handling multi-session authentication. However, there is a notable limitation in the availability of datasets for multi-session studies. Our investigation focuses on comparing single-session and multi-session authentication to understand the performance gap between these two approaches. We expected an increase in EER for multi-session authentication, as it represents a more complex challenge. This increase in difficulty is attributed to additional noise factors inherent in multi-session settings. Additionally, variations in EEG electrode placement across sessions and possible changes in brain states add to the complexity, making multi-session authentication a more challenging task compared to single-session scenarios.

\begin{figure}[H]
    \centering
    \includegraphics[width=1\linewidth]{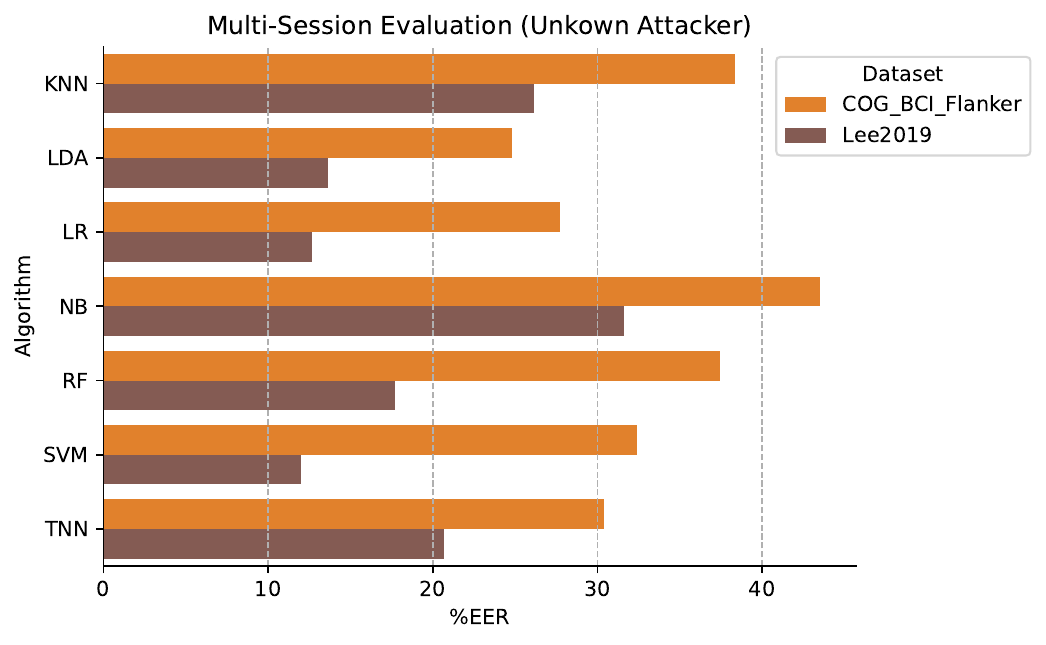}
    \caption{Comparing the effectiveness of shallow classifiers and TNN for authentication against unknown attackers in two multi-session datasets.}
    \label{fig:Multi_Session}
\end{figure}

Figure \ref{fig:Multi_Session} presents the results for multi-session authentication, where the results indicate a significant increase in EER compared to the single-session results shown in Figure \ref{fig:Session_Evaluation}. Notably, LDA and LR, which were not among the best performers in the single-session scenario, show more promising results in the multi-session context. These findings highlight the need for further research focused on multi-session scenarios to better understand and improve authentication performance in these more complex settings. The average EERs of 21\% and 30\% in multi-session scenarios indicate limitations in the practical implementation of a brainwave biometric authentication system, especially considering the False Rejection Rate (FNMR) at False Acceptance Rates (FMR) of 1\%, 0.1\%, and 0.01\%. These rates were respectively 72.61\%, 86.70\%, 89.50\% for the Lee2019 dataset and 84.43\%, 94.05\%, 94.09\% for the COG\_BCI\_Flanker dataset (Table \ref{tab:fnmr}), suggesting the need for improved models to achieve lower EERs for real-world applications.

The lower performance of the deep learning approach compared to shallow classifiers in multi-session scenarios may result from the limited amount of data available for training feature extraction models. These models face complex issues across different sessions, such as changes in hair length, emotional states, and electrode positions, which complicate effective feature extraction with limited data. In contrast, classical feature extraction methods like PSD or AR require less data by relying on expert knowledge and do not need data for learning features. However, with more data, we expect twin neural network models to improve and potentially outperform shallow classifiers.

\subsection{Shallow Classifiers vs Similarity-Based Approach}
\label{sec:shallow}
Comparing the training of a general model to extract identity-specific features with training individual models for each subject based on traditional feature extraction represents two different approaches. We utilized a twin neural network (TNN) as a representative of the similarity-based approach. We anticipated that the similarity approach would yield better results, as it actively seeks to correlate features from raw brainwaves, unlike shallow classifiers which attempt to discern differences between PSD and AR features of a subject and attacker's samples.

The results revealed that the TNN approach, along with RF, was among the top-performing models in single-session results (Table \ref{tab:fnmr}). In multi-session results, despite having performance close to RF, other classifiers like LDA and LR showed better performance (Figure \ref{fig:Multi_Session}). This suggests the necessity for either more comprehensive data or an enhanced deep learning strategy to surpass traditional classifiers that utilize straightforward feature extraction methods.

\begin{table*}[p]
\captionsetup{font=small} 
    \caption{\small{Comparative Analysis of Average EER and FNMR at 1$\%$, 0.1$\%$, and 0.01$\%$ FMR thresholds across nine Datasets. The evaluation encompasses both single-session and multi-session Schemes, with a focus on classifiers' performance in the unknown attacker scenario. Values are presented in percentages.}}
    \label{tab:Table 3}
    \small
    \begin{tabular}{@{}p{2.2cm} @{\hspace{0.5cm}} p{2.0cm} @{\hspace{0.5cm}} p{2.2cm} @{\hspace{0.9cm}}| *{9}{p{1.0cm}}|p{1.0cm}}
        \toprule
        \multirow{2}{*}{\textbf{Dataset}} & \multirow{2}{*}{\textbf{Evaluation}} & \multirow{2}{*}{\textbf{Metric}} & \multirow{2}{*}{\textbf{KNN}} & \multirow{2}{*}{\textbf{LDA}} & \multirow{2}{*}{\textbf{LR}} & \multirow{2}{*}{\textbf{NB}} & \multirow{2}{*}{\textbf{RF}} & \multirow{2}{*}{\textbf{SVM}} & \multirow{2}{*}{\textbf{TNN}}\\
        & & & & & & & & & \\
        \midrule
        & & $\%$EER & 2.44 & 12.92 & 5.14 & 4.45 & 0.46 & 0.74 & \textbf{0.43}\\
        BrainInvaders15a & Single-Session & \mbox{FNMR at 1\% FMR} & 19.87 & 71.26 & 43.39 & 31.55 & \textbf{0.69} & 2.06 & 2.12\\
        & & \mbox{FNMR at 0.1\% FMR} & 34.62 & 83.22 & 58.55 & 38.77	 & \textbf{1.47} & 4.54	 & 3.70\\
        & & \mbox{FNMR at 0.01$\%$ FMR} & 46.56 & 89.52 & 71.30 & 45.55 & \textbf{2.06} & 6.01 & 5.11\\
        \midrule
    
        & & $\%$EER & 5.72 & 4.75 & 3.41 & 10.16 & 0.90 & 2.73	 & \textbf{0.81}\\
        ERPCORE\_N400 & Single-Session & \mbox{FNMR at 1\% FMR} & 22.70 & 35.83 & 25.67 & 76.06 & 1.96 & 7.54 & \textbf{1.17}\\
        & & \mbox{FNMR at 0.1$\%$ FMR} & 45.11 & 72.76 & 68.53 & 94.55 & 4.95 & 15.87 & \textbf{3.96}\\
        & & \mbox{FNMR at 0.01$\%$ FMR} & 52.66 & 75.26 & 68.54 & 97.80 & 5.11 & 15.87 & \textbf{3.96}\\
        \midrule
       
        & & $\%$EER & 6.87	 & 5.73 & 3.45 & 9.11 & \textbf{1.37} & 3.68 & 1.53\\
        ERPCORE\_P300 & Single-Session & \mbox{FNMR at 1\% FMR} & 26.65 & 39.28 & 30.03 & 59.55 & 4.36	 & 10.47 & \textbf{4.30}\\
        & & \mbox{FNMR at 0.1$\%$ FMR} & 46.59 & 65.97 & 55.84 & 84.84 & \textbf{7.67} & 19.11 & 9.63\\
        & & \mbox{FNMR at 0.01$\%$ FMR} & 49.88 & 67.50 & 55.88 & 88.83 & \textbf{7.83} & 19.11 & 9.63\\
        
        \midrule
        & & $\%$EER & 3.28 & 1.58 & 0.84 & 5.89 & 0.36 & 0.55 & \textbf{0.03}\\
        Huebner\_LLP & Single-Session & \mbox{FNMR at 1\% FMR} & 26.48 & 8.41 & 5.06 & 34.01 & 0.31 & 0.51 & \textbf{0.02}\\
        & & \mbox{FNMR at 0.1$\%$ FMR} & 54.90 & 25.74 & 16.71 & 51.88 & 0.78	 & 0.79	 & \textbf{0.02}\\
        & & \mbox{FNMR at 0.01$\%$ FMR} & 74.21 & 54.26 & 32.76 & 63.84 & 1.38 & 1.19 & \textbf{0.05}\\
        
        \midrule
        & & $\%$EER & 8.56	 & 4.76 & 3.50 & 20.05 & \textbf{3.78} & 4.04 & 4.76\\
        Mantegna2019 & Single-Session & \mbox{FNMR at 1\% FMR} & 31.26 & 38.30 & 28.26 & 86.57 & \textbf{9.88} & 11.04 & 19.61\\
        & & \mbox{FNMR at 0.1$\%$ FMR} & 58.51 & 71.97 & 63.33 & 96.54 & \textbf{22.65} & 23.88 & 46.53\\
        & & \mbox{FNMR at 0.01$\%$ FMR} & 66.94 & 73.53 & 63.42 & 98.23 & \textbf{22.93} & 23.88 & 46.53\\
         
        \midrule

        & & $\%$EER & 4.22 & 6.31 & 7.73 & 9.11 & 0.38 & 0.79 & \textbf{0.04}\\
        Sosulski2019 & Single-Session & \mbox{FNMR at 1\% FMR} & 21.60 & 33.08 & 28.77 & 50.93 & 0.42 & 1.59 & \textbf{0.01}\\
        & & \mbox{FNMR at 0.1$\%$ FMR} & 45.47 & 37.98 & 35.01 & 63.00 & 0.68 & 3.34 & \textbf{0.05}\\
        & & \mbox{FNMR at 0.01$\%$ FMR} & 65.65 & 45.01 & 41.66 & 76.08 & 1.09 & 4.94 & \textbf{0.09}\\
        
        \midrule
         & & $\%$EER & 4.46 & 5.33 & 2.86 & 12.89 & 1.27 & 1.96 & \textbf{1.02}\\
        Won2022 & Single-Session & \mbox{FNMR at 1\% FMR} & 18.37 & 45.10 & 27.72 & 84.48	 & 3.38 & 4.93 & \textbf{2.07}\\
        & & \mbox{FNMR at 0.1$\%$ FMR} & 44.87 & 76.98 & 69.14 & 96.63 & 8.88 & 13.68 & \textbf{8.66}\\
        & & \mbox{FNMR at 0.01$\%$ FMR} & 67.33 & 89.96 & 84.89 & 98.84 & 14.82 & 20.06 & \textbf{14.31}\\
        
        \midrule
        & & $\%$EER & 7.69 & 5.05 & 3.82 & 9.62 & \textbf{3.16} & 4.18 & 11.00\\
        COG\_BCI\_Flanker & Single-Session & \mbox{FNMR at 1\% FMR} & 32.11 & 36.73 & 27.70 & 56.51 & \textbf{11.30} & 14.25 & 51.74\\
        & & \mbox{FNMR at 0.1$\%$ FMR} & 52.84 & 57.97 & 49.48 & 70.61 & \textbf{16.97} & 21.00 & 62.66\\
        & & \mbox{FNMR at 0.01$\%$ FMR} & 57.13 & 59.91 & 49.51 & 73.15 & \textbf{17.23} & 21.00 & 62.66\\
        \\
        & & $\%$EER & 38.35 & 24.80 & \textbf{27.75} & 43.54  & 37.44 & 32.39 & 30.37\\
        COG\_BCI\_Flanker & Multi-Session & \mbox{FNMR at 1\% FMR} & 82.36 & 82.63 & \textbf{79.68} & 92.20 & 83.64 & 76.98 & 84.43\\
        & & \mbox{FNMR at 0.1$\%$ FMR} & 92.44 & 90.18 & \textbf{88.06} & 96.65 & 88.68 & 84.08 & 94.05\\
        & & \mbox{FNMR at 0.01$\%$ FMR} & 94.11 & 90.29 & \textbf{88.06} & 97.11 & 88.87 & 84.08 & 94.09\\
        
        \midrule
        & & $\%$EER & 6.93	 & 3.47 & 1.48 & 20.85 & 2.14 & 2.17 & \textbf{1.39}\\
        Lee2019 & Single-Session & \mbox{FNMR at 1\% FMR} & 20.71	 & 26.65 & 10.79 & 90.89 & 4.90 & 4.56 & \textbf{3.53}\\
        & & \mbox{FNMR at 0.1$\%$ FMR} & 42.90 & 58.43 & 38.52 & 97.18 & \textbf{11.11} & 11.63 & 13.56\\
        & & \mbox{FNMR at 0.01$\%$ FMR} & 58.08 & 71.00 & 52.93 & 97.96 & \textbf{14.06} & 15.70 & 19.39\\
        \\
        & & $\%$EER & 26.13 & 13.63 & 12.65 & 31.64 & 17.73 & \textbf{11.99} & 20.68\\
        Lee2019 & Multi-Session & \mbox{FNMR at 1\% FMR} & 63.67 & 66.56 & 58.93 & 94.84 & 55.84 & \textbf{45.57} & 72.61\\
        & & \mbox{FNMR at 0.1$\%$ FMR} & 82.23 & 85.21 & 84.53 & 99.50 & 70.39 & \textbf{62.54} & 86.70\\
        & & \mbox{FNMR at 0.01$\%$ FMR} & 89.96 & 88.16 & 89.79 & 99.95 & 74.92 & \textbf{67.81} & 89.50\\
        \bottomrule
    \end{tabular}

    \label{tab:fnmr}
\end{table*}

\subsection{Comparison with Related Work}
The comparison of our results will initially focus on the results published by Arias et al. \cite{arias2021inexpensive} and Fallahi et al. \cite{fallahi2023brainnet}. This is due to the establishment of our benchmark, which is based on their implementations. Moreover, both studies utilized the ERPCORE~\cite{erpcore} P300 and N400 datasets, similar to our approach. Arias et al. \cite{arias2021inexpensive} investigate the performance of different shallow classifiers within single-session brainwave authentication. The results indicate 1.9\% EER for N400 and 3\% EER for P300 paradigm with a sample rejection of 120 µV and RF classifier, we observe the same results where we achieved  2.9\% EER for P300 and 1.13\% EER to N400 paradigm with the sample rejection of 100 µV and RF classifier. Also, Fallahi et al. \cite{fallahi2023brainnet} acquired 2.01\% and 1.37\% respectively for P300 and N400 ERPCORE and 0.14 for P300:bi2015a dataset~\cite{bi2015a}, where we obtained \%1.53,  0.83\% , and 0.43 respectively. As can be observed the results are fairly similar and trends are the same. The small difference in results could be from a different random seed, version of libraries, and some evaluation parameters like epoch rejection rate for shallow classifiers and number of epochs for twin neural networks. To enhance the reproducibility of our result, we provided a docker container in the GitHub of NeuroIDBench. 

\textbf{Preprocessing:} Our results about the impact of sample duration corroborate prior research on shallow classifiers, which indicates that longer sample durations can result in reduced EER or increased accuracy \cite{arias2021inexpensive,tran2019eeg,carrion2019method,suppiah2018biometric}. However, the current version of the twin neural network requires modification to effectively utilize longer samples.

\textbf{Feature Extraction:} The results of our feature extraction analysis demonstrate that PSD features notably outperform AR features, corroborating findings from Huang et al.~\cite{huang2022m3cv}. However, there is a lack of consensus in the literature regarding the optimal order for AR features. Studies by Arias et al. \cite{arias2023performance} and Brigham \cite{brigham2010subject} suggest that lower AR orders yield better results, whereas research by Kaewwit et al.~\cite{kaewwit2017high} and Zhang et al. \cite{zhang2017classification} indicates improved outcomes with higher orders. Our findings align with the former, showing that lower-order AR features are more effective in 8 out of 9 datasets analyzed. 

\textbf{Known and Unkown attacker:} In exploring scenarios involving known and unknown attackers, Wu et al. \cite{wu2018eeg} conducted research but found no notable performance decline in the unknown attacker scenario, and even reported an improvement. Similarly, Panzino et al. \cite{panzino2022eeg} studied the same issue and observed only a 5\% decrease in accuracy. Contrarily, Arias et al.\cite{arias2023performance} and Fallahi et al.\cite{fallahi2023brainnet} reported an increase in Equal Error Rate (EER) ranging from 10\% to 17.5 times. Our results indicate that the unknown attacker scenario is consistently more effective than the known attacker scenario, although the extent of this difference varies across different datasets and models. An average calculation across nine datasets and seven models shows that the EER for unknown attackers is 60.2\% higher than known attacker scenarios.

\textbf{Single and Multi-session:} Multi-session analysis is crucial for the practical application of brainwave authentication, yet it presents heightened challenges compared to single-session scenarios. In multi-session environments, the availability of public datasets is notably scarcer, and the practice of open-source sharing becomes more constrained. Results from various studies, such as Maiorana \cite{maiorana2021learning}, demonstrate reasonable EER as low as 4.8\%. Seha et al. \cite{seha2019new} also reported a marginal difference of less than 0.5\% EER between single and multi-session scenarios under specific feature extraction methods and LDA classifiers. Furthermore, Wu et al. \cite{wu2018eeg} observed a 1.3\% improvement in false rejection rate performance in the second session compared to the first session in single-session scenarios. In contrast, our research reveals a significant disparity between single-session and multi-session results, with our best outcome showing an 11.99\% EER. This finding aligns with Huang et al. \cite{huang2022m3cv}, who also reported a substantial difference between within-session and cross-session evaluations. However, it is important to note that the studies by Maiorana \cite{maiorana2021learning}, Seha et al. \cite{seha2019new}, and Wu et al. \cite{wu2018eeg} do not provide public datasets or open-source code for other researchers to utilize and build upon. Therefore, there is an urgent need for more open-source and publicly available datasets in multi-session studies in this field. In fact, understanding the exact state of this challenge is crucial to determine what steps are necessary to advance the field.

\textbf{Shallow Classifiers vs Similarity-Based Approach:}
The current trend in biometric authentication leans towards deep learning approaches. We observed the same trend in brainwave authentication where several papers based on TNN \cite{fallahi2023brainnet,maiorana2021learning,schons2018convolutional} or CNN \cite{bidgoly2022towards} to improve authentication.
However, a comparison between shallow classifiers and deep learning methods is essential. This is because the scarcity of data can hinder the training of deep learning networks, particularly in managing the high variation in brainwave data. Several studies~\cite{debie2021session,fallahi2023brainnet} demonstrate that deep learning approaches can outperform shallow classifiers. However, our large-scale benchmark emphasizes, despite the impressive results of TNN, they do not consistently surpass shallow classifies, Notably, in multi-session scenarios, TNNs underperform,  potentially due to insufficient data to accommodate the high variability between sessions.

\section{Limitations}
The two main limitations observed while conducting research to develop our benchmark are the shortage of publicly available datasets and the evolving ethical and privacy concerns in the deployment of EEG-based biometric systems. These challenges underscore critical areas for future research and development to enhance the reliability and ethical assurance of EEG authentication.

\textbf{Publicly Available Datasets:} Currently available public datasets are not designed for authentication purposes. They often have a low number of subjects, usually fewer than 70, and mostly provide data from a single session, making it difficult to evaluate the robustness of EEG authentication over time (Table \ref{tab:erp_datasets}). Additionally, most datasets use medical-grade devices for data collection, complicating performance estimation in real-world scenarios. Therefore, there is a need to collect authentication datasets that include a larger number of subjects (over 500), with at least three sessions, and various environmental conditions, such as variations in noise, lighting, hair length, and emotional states. This will enable proper training and evaluation of models and allow for comparisons with well-known biometrics such as face recognition models.

\textbf{Ethical and Privacy Implications:} As EEG authentication research advances toward practical applications, increasing attention is being paid to the ethical aspects of collecting and using brain data for authentication. 
 Previous studies have looked into privacy concerns from researchers~\cite{holler2018eeg,wang2022polycosgraph} and users~\cite{rose2023overcoming,fallahi2024usability}, identifying the need for protecting these data. The main issue is that EEG can reveal highly sensitive  personal information, including  emotional states~\cite{wang2014emotional}, medical conditions~\cite{sanchez2021impact}, attention levels~\cite{hassan2020human}, and gender~\cite{niu2024gender}. Therefore, collecting brain data raises the risk of harmful data breaches and introduces new possibilities for abuse or misuse. For example, an honest-but-curious authentication provider could perform unauthorized behavioral monitoring or exploitation of brain reactions for commercial purposes. This potential threat is aggravated by the lack of user awareness regarding the sensitivity of brain data when using commercial BCIs \cite{kablo2023privacy}. Beyond privacy-related concerns, there is a risk of bias in deployed EEG-based authentication systems if not trained with a diverse set of people. This type of issue has been observed in face recognition software that performs poorly when used by subjects who belong to underrepresented groups \cite{buolamwini2018gender}.

 To address these concerns, we recommend the following general guidelines for ethical practice. In terms of data collection for research, researchers must adhere to the ethical standards established in the Menlo Report \cite{bailey2012menlo} 
 and inform subjects about potential risks through consent forms. For real-world EEG authentication, it is crucial to develop robust template protection methods to prevent authentication providers from accessing raw EEG data, protecting against potential leaks and inferences. Systems should be trained to ensure fairness and reduce algorithmic bias, which entails the collection of inclusive datasets that capture the diversity of all potential users.  Similarly, broad studies on the social acceptance and diverse user needs should be conducted to guide a responsible human-centered development of EEG biometrics. Finally, it is important to implement effective transparency mechanisms beyond unusable privacy policies to communicate to BCI users what the collected data and privacy risks are.

\section{Conclusion and Future Work}
\label{sec:sensors}
In conclusion, NeuroIDBench is presented as a comprehensive benchmarking tool for EEG authentication, particularly supporting ERP-based methods. As an open-source platform, it currently supports 9 datasets and 7 classifiers. It can easily be adapted to include new datasets and classifiers, allowing for the investigation of different methodological research questions. Our results highlight that epoch rejection does not significantly affect outcomes. PSD combined with AR order 1 emerges as a recommended default feature set for shallow classifiers. Known attacker evaluation can misleadingly indicate lower EER. Multi-session evaluation is significantly more challenging than single-session evaluation. Moreover, it appears that due to the limited availability of data, deep learning approaches do not consistently outperform shallow classifiers in brainwave-based authentication. 

We encourage researchers to contribute to this open-source project, aiding in the development of more practical and diverse EEG-based authentication systems.

\section*{Acknowledgements}
This work was funded by the Topic Engineering Secure Systems of the Helmholtz Association (HGF) and supported by KASTEL Security Research Labs, Karlsruhe and Germany’s Excellence Strategy (EXC 2050/1 ‘CeTI’; ID 390696704). We thank the textician of KASTEL Security Research Labs for assistance and support in the research communication.

\bibliographystyle{elsarticle-num} 
\bibliography{bibliography}

\clearpage
\appendix
\section{Receiver Operating Characteristic (ROC) Curves}

 ROC curves play a pivotal role in assessing the performance of biometric systems like brainwave authentication. These curves provide a graphical representation of the system's ability to distinguish between genuine users and impostors by plotting FMR against 1-FNMR across various threshold values. 
 In brainwave authentication systems, ROC curves help in determining the optimal threshold for achieving the desired balance between sensitivity and specificity, thereby ensuring the system's effectiveness in accurately identifying authorized users while minimizing the risk of false positives and false negatives.

\subsection{Single Session Evaluation (Unknown Attacker Scenario)}
\smallskip

\smallskip

\begin{figure}[H]
    \centering
    \includegraphics[width=0.75\linewidth]{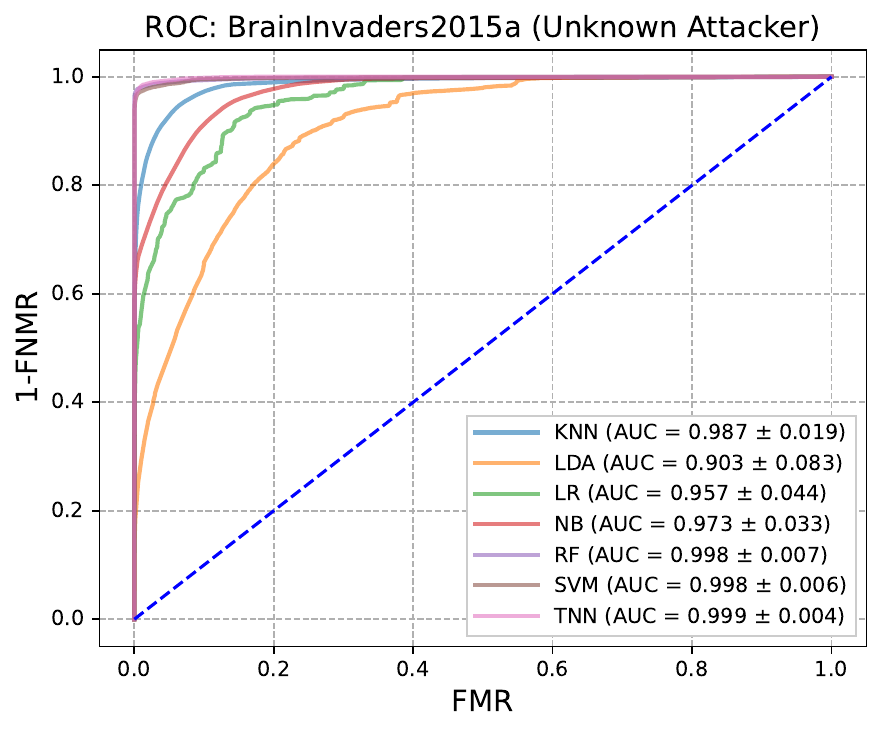}
    \caption{ROC for dataset BrainInvaders15a in Single Session Evaluation under unknown attacker Scenario.}
\end{figure}

\begin{figure}[H]
    \centering
    \includegraphics[width=0.75\linewidth]{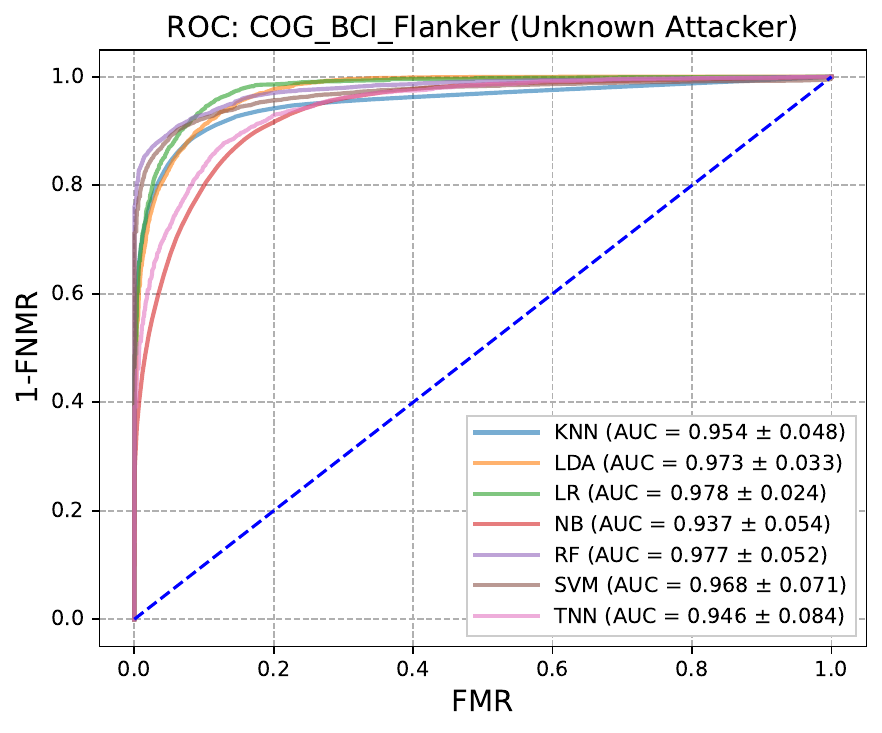}
    \caption{ROC for dataset COG\_BCI\_Flanker in Single Session Evaluation under unknown attacker Scenario.}
\end{figure}

\begin{figure}[H]
    \centering
    \includegraphics[width=0.75\linewidth]{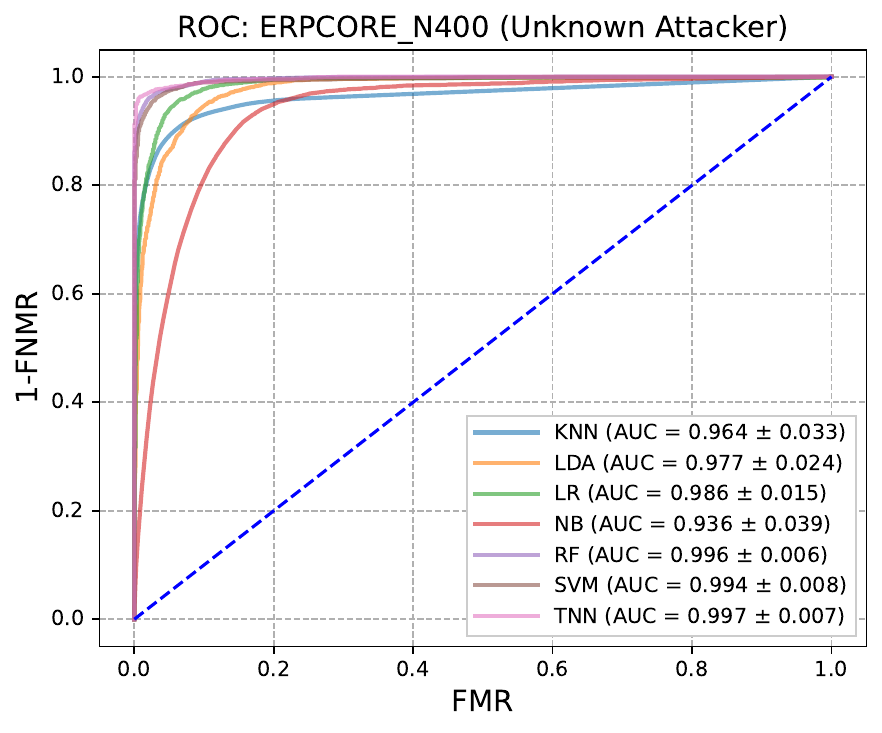}
    \caption{ROC for dataset ERPCORE\_N400 in Single Session Evaluation under unknown attacker Scenario.}
\end{figure}

\begin{figure}[H]
    \centering
    \includegraphics[width=0.75\linewidth]{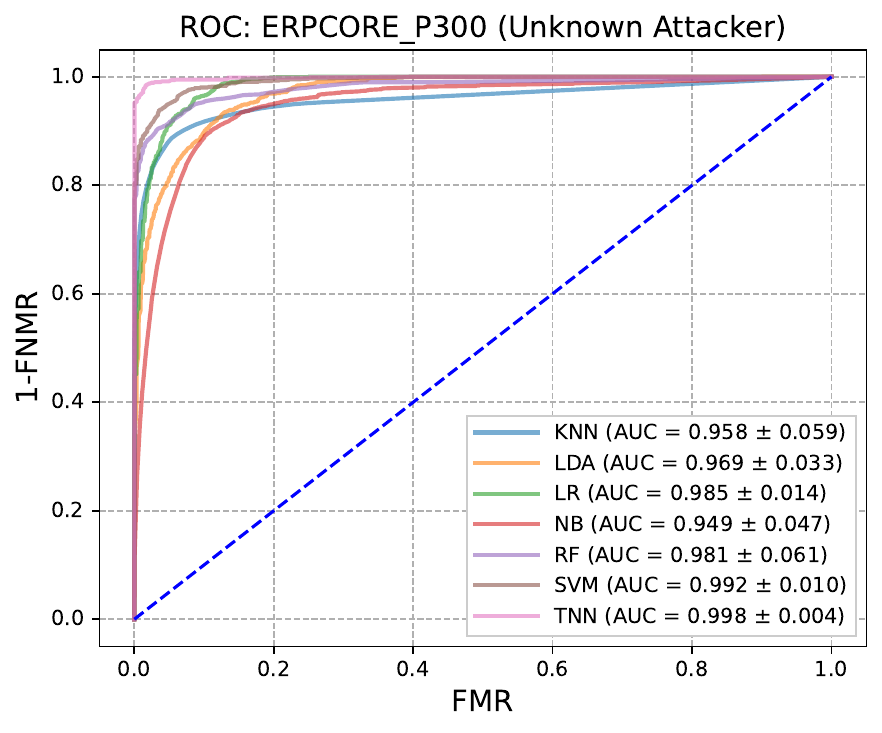}
    \caption{ROC for dataset ERPCORE\_P300 in Single Session Evaluation under unknown attacker Scenario.}
\end{figure}

\begin{figure}[H]
    \centering
    \includegraphics[width=0.75\linewidth]{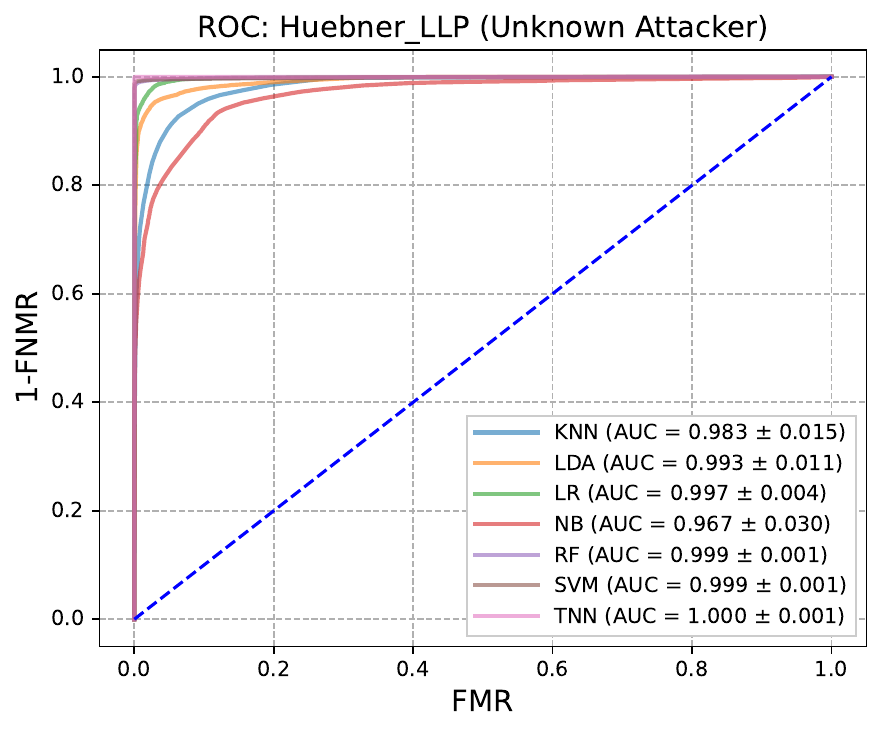}
    \caption{ROC for dataset Huebner\_LLP in Single Session Evaluation under unknown attacker Scenario.}
\end{figure}

\begin{figure}[H]
    \centering
    \includegraphics[width=0.75\linewidth]{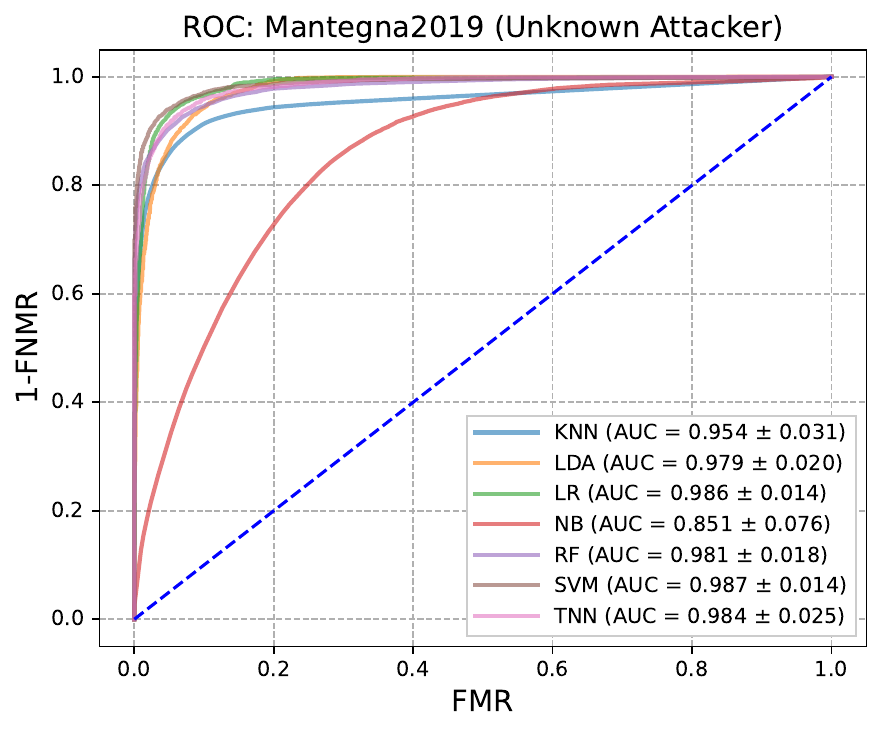}
    \caption{ROC for dataset Mategna2019 in Single Session Evaluation under unknown attacker Scenario.}
\end{figure}

\begin{figure}[H]
    \centering
    \includegraphics[width=0.75\linewidth]{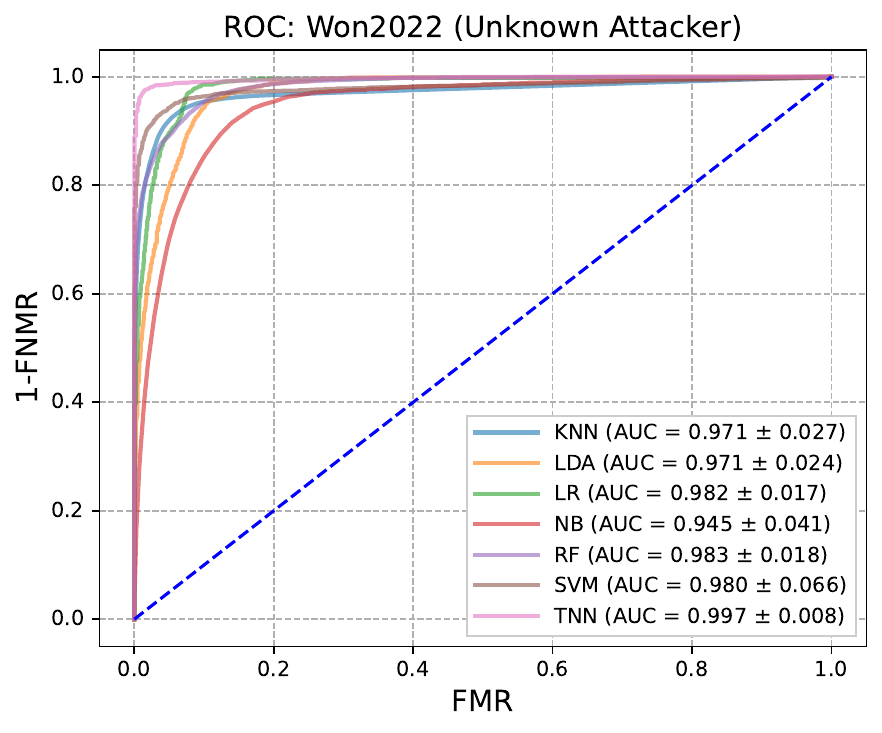}
    \caption{ROC for dataset Won2022 in Single Session Evaluation under unknown attacker Scenario.}
\end{figure}

\begin{figure}[H]
    \centering
    \includegraphics[width=0.75\linewidth]{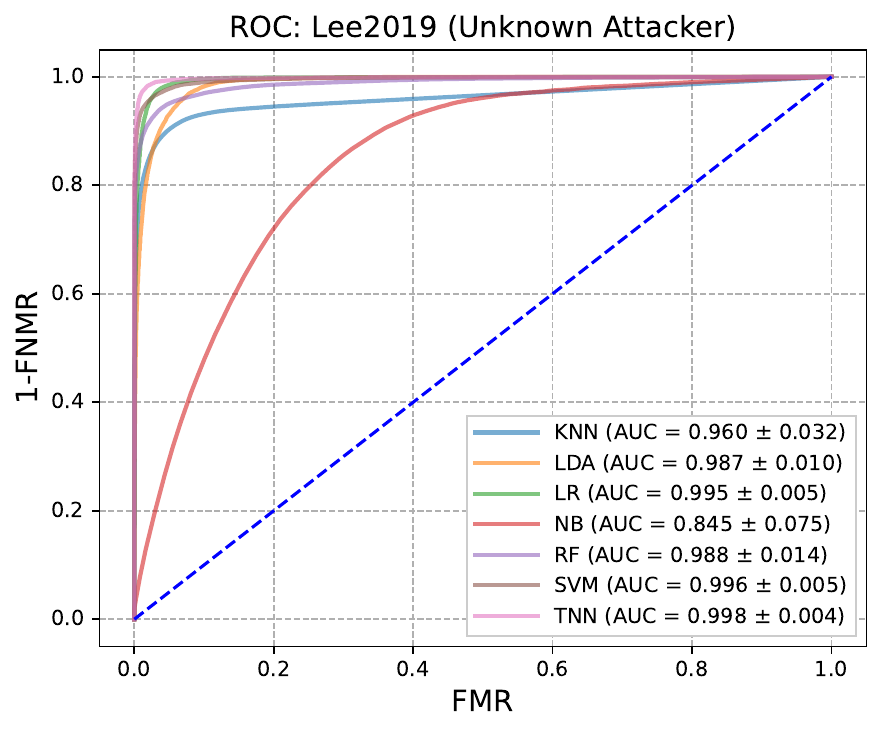}
    \caption{ROC for dataset Lee2019 in Single Session Evaluation under unknown attacker Scenario.}
\end{figure}

\begin{figure}[H]
    \centering
    \includegraphics[width=0.75\linewidth]{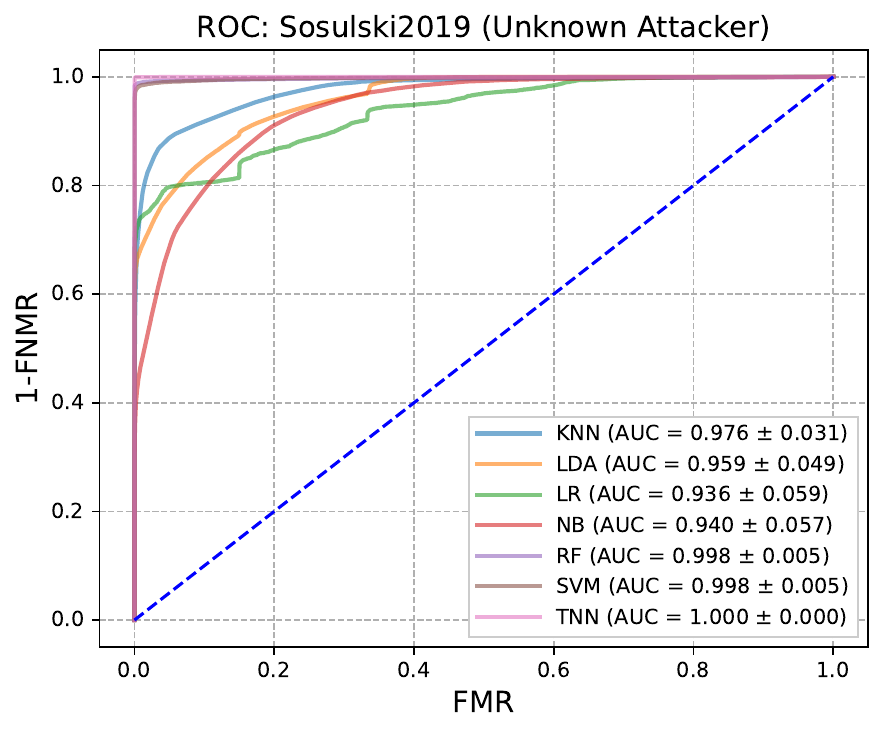}
    \caption{ROC for dataset Lee2019 in Single Session Evaluation under unknown attacker Scenario.}
\end{figure}

\subsection{Multi Session Evaluation (Unknown Attacker Scenario)}

\begin{figure}[H]
    \centering
    \includegraphics[width=0.75\linewidth]{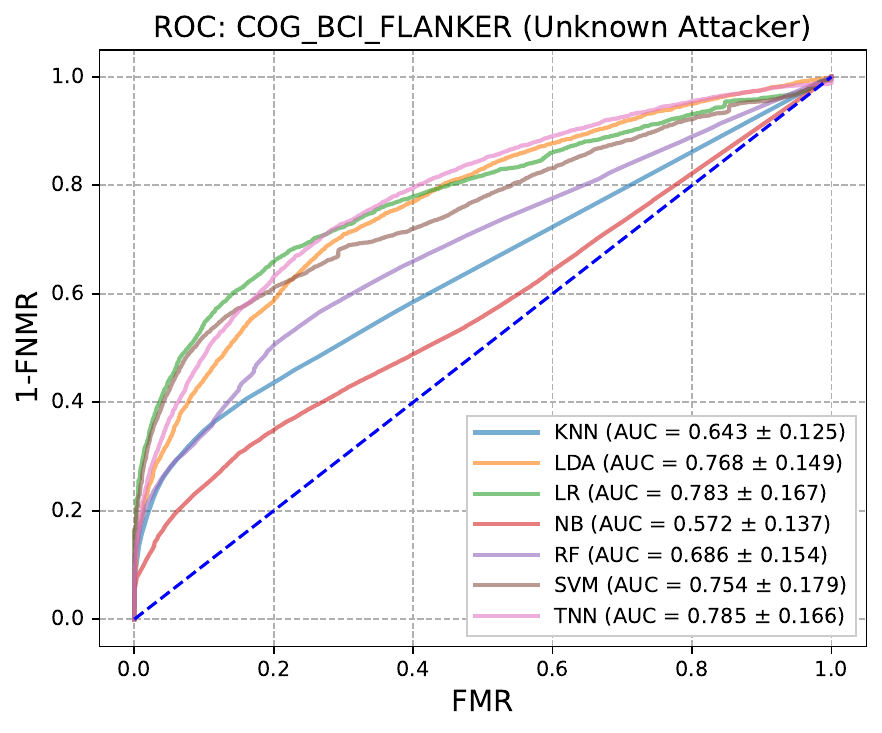}
    \caption{ROC for dataset COG\_BCI\_Flanker in Multi Session Evaluation under unknown attacker Scenario.}
\end{figure}

\begin{figure}[H]
    \centering
    \includegraphics[width=0.75\linewidth]{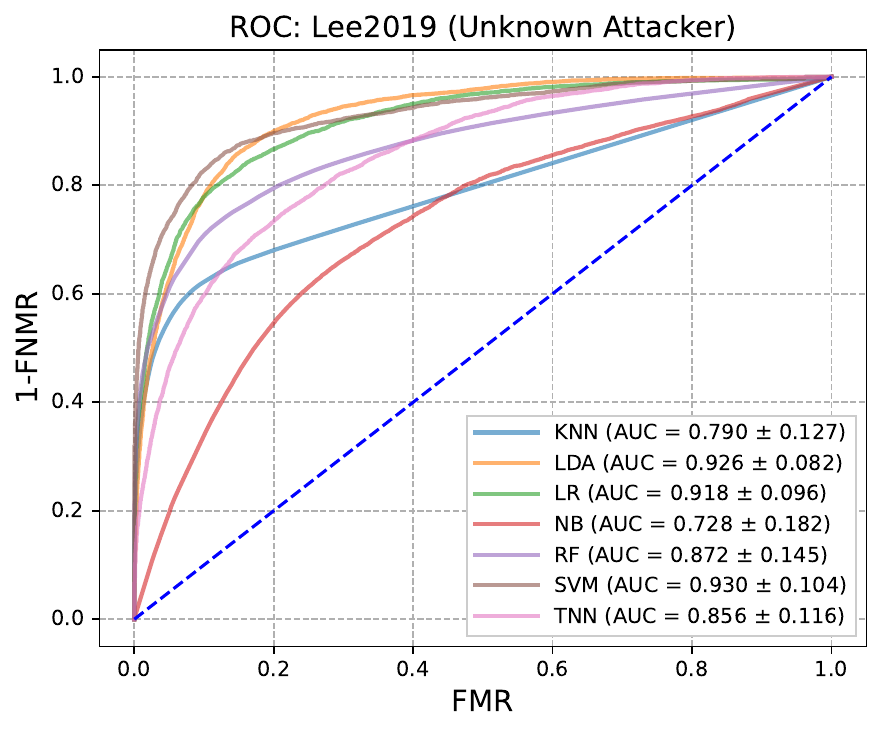}
    \caption{ROC for dataset Lee2019 in Multi Session Evaluation under unknown attacker Scenario.}
\end{figure}

\section{Installation and Running the tool}
\label{Appendix Installation}
The benchmarking tool is available on our Github repository 
\footnote{https://github.com/Avichaurasia/NeuroIDBench.git}. Developed in Python, the tool leverages various statistical and machine-learning libraries, necessitating the establishment of a Python environment as a preliminary step. The subsequent procedures delineate the steps to set up the operational envirmoment and replicate outcomes utilizing the NeuroIDBench tool. 

\subsection{Establishing the Running Environment}
The first step after extracting the tool from Github is to set up the running environment first. An optimal approach that involves creating a Python virtual environment dedicated to the framework, achieved through the creation of a Conda environment. This environment encapsulates all Python dependencies essential for executing the tool seamlessly. Alternatively, users may opt to containerize the Python package using Docker, thereby providing an isolated environment that encapsulates all dependencies required to execute the tool within the container. This approach ensures reproducibility and facilitates the seamless deployment of the benchmarking tool.
\\
\\
\textbf{Setting up the Conda Environment}
\\
\\
To successfully run this Python project, creating and configuring a suitable Python environment is essential. The following steps need to be followed to set up the environment:
\begin{enumerate}
    \item \textbf{Python Installation}: 
    Initially, verifying the presence of Python in our operating system is crucial. If Python is not pre-installed, it can be acquired straight from the official Python\footnote{https://https://www.python.org/downloads/} website. Subsequently, we adhere to the installation instructions outlined in the website, customized to suit our operating system. If Anaconda has been successfully installed, it is noteworthy to mention that the Anaconda installation often includes the Python programming language. If the action above is taken, it is possible to go to the subsequent stage.
\\

    \item \textbf{Anaconda Installation (if needed)}: 
    Suppose Anaconda is not currently installed, and it is desired to utilize it to manage Python environments. In that case, it is possible to get the software by downloading it from the official website, which can be accessed at Anaconda \footnote{https://www.anaconda.com/products/individual} website. Installation may be accomplished according to the instructions tailored to each operating system. Anaconda provides a user-friendly method for creating and administrating virtual environments through the use of Conda. This specific characteristic has significant value in data science and scientific computing initiatives.
\\

    \item \textbf{Virtual Environment Creation}: It is advisable to establish a virtual environment to segregate the dependencies of this specific project from other Python packages installed on our system. Virtual environments help maintain clean and distinct Python environments for individual projects. To create a virtual environment, follow these steps: 
    \begin{itemize}
        \item \textbf{Navigate to Project Directory}: To begin, we access the terminal or command prompt and proceed to the project's root directory by utilizing the \textit{cd} command. 

        For example: cd /path/to/project
        \\
        
        \item \textbf{Environment Configuration File}: Check if the project includes an environment configuration file. This file is typically named \textit{environment.yml} or \textit{requirements.txt} and lists the required Python packages and their versions.
        \\
        
        \item \textbf{Create Virtual Environment}:Subsequently, the requisite command is executed to generate a virtual environment by using the configuration file. An example of a command that may be used for Conda environments is the use of a \textit{environment.yml} file:
        
        For example: conda env create -f environment.yml
        \\
        
        we can also utilize \textit{requirement.txt} to create the virtual environment by using the pip command:
        
        python -m venv venv  
        
        source venv/bin/activate  
        
        pip install -r requirements.txt
        \\
        
    \end{itemize}
    
    \item \textbf{Activate the Virtual Environment}: After the virtual environment has been established, proceed to activate it. Activation is a crucial process that guarantees using an isolated environment and its corresponding dependencies in our project. To activate the conda environment, it is necessary to utilize the proper command according to the operating system in use:
        
    For example: conda activate master\_thesis (for MacOS/Linux)
\end{enumerate}
\
\
 \textbf{Setting up Docker Container}
\\

To containerize a project, we typically use a tool like Docker. Here's a step-by-step guide on how to do it:

\begin{enumerate}
    \item Install Docker on the local machine if we have not already. Install docker from the official Docker \footnote{https://www.docker.com/} website.
\\

    \item Create a \textbf{dockerfile} in the root directory of the framework. This file will contain the instructions for Docker to build our container.
\\
\item Get the Python docker image python:3.8-slim-buster from Docker Hub \footnote{https://hub.docker.com/}.
\\

\item Edit the dockerfile with the following steps: 
\\

FROM python:3.8-slim-buster
\\
WORKDIR /app
\\
ADD . /app
\\
COPY requirements.txt ./
\\
RUN pip install --upgrade pip
\\
RUN apt-get update \&\& apt-get install -y $\backslash$
\\
     build-essential $\backslash$
    \\
    libblas-dev $\backslash$
    \\
    liblapack-dev $\backslash$
    \\
    gfortran $\backslash$
    \\
    libhdf5-dev $\backslash$
    \\
    cython $\backslash$
    \\
    pkg-config $\backslash$
\\
RUN pip install --no-binary=h5py h5py
\\
RUN pip install -r new\_requirements.txt
\\
WORKDIR /app/neuroIDBench
\\
CMD ["python", "run.py"]
\\
\item Build the docker image from the dockerfile, using the following command: 
\\
docker build -t dockerfile .
\\
\end{enumerate}

\subsection{YAML configurations}
Once the Python environment or Docker container is developed, we need to edit the YAML configuration files that can be found as \textit{single\_dataset.yml} under the configuration\_files folder. Following are some of the examples of the YAML configuration files that can be utilized to perform automated benchmarking and reproduce the results for the single dataset.   
\begin{lstlisting}[basicstyle=\ttfamily\small, caption={Benchmarking pipeline using the dataset's default parameters and auto-regressive features with SVM classification}]

name: "BrainInvaders2015a"

dataset:
  - name: BrainInvaders2015a
    from: neuroIDBench.datasets
    
pipelines:
  "AR+SVM":
    - name: AutoRegressive
      from: neuroIDBench.featureExtraction
     
    - name: SVC
      from: sklearn.svm
      parameters:
        kernel: 'rbf'
        class_weight: "balanced"
        probability: True
\end{lstlisting}

\begin{lstlisting}[basicstyle=\ttfamily\small, caption={Benchmarking pipeline using dataset's parameters and Auto Regressive order with SVM classification}]

name: "BrainInvaders2015a"

dataset:
  - name: BrainInvaders2015a
    from: neuroIDBench.datasets
    parameters:
      subjects: 10
      interval: [-0.1, 0.9]
      rejection_threshold: 200

pipelines:
  "AR+SVM":
    - name: AutoRegressive
      from: neuroIDBench.featureExtraction
      parameters:
        order: 5
      
    - name: SVC
      from: sklearn.svm
      parameters:
        kernel: 'rbf'
        class_weight: "balanced"
        probability: True
\end{lstlisting}

\begin{lstlisting}[basicstyle=\ttfamily\small, caption={Benchamrking pipeline for dataset BrainInvaders15a with AR and PSD features with classifier SVM}]

name: "BrainInvaders2015a"

dataset:
  - name: BrainInvaders2015a
    from: neuroIDBench.datasets
    parameters:
      subjects: 10
      interval: [-0.1, 0.9]
      rejection_threshold: 200
      
pipelines:
  "AR+PSD+SVM":
    - name: AutoRegressive
      from: neuroIDBench.featureExtraction
      parameters:
        order: 5
      
    - name: PowerSpectralDensity
      from: neuroIDBench.featureExtraction
      
    - name: SVC
      from: sklearn.svm
      parameters:
        kernel: 'rbf'
        class_weight: "balanced"
        probability: True
\end{lstlisting}

\begin{lstlisting}[basicstyle=\ttfamily\small, caption={Benchamrking pipeline for dataset BrainInvaders15a with Twin NN}]
name: "BrainInvaders2015a"

dataset: 
  - name: BrainInvaders2015a
    from: neuroIDBench.datasets
    parameters: 
      subjects: 10
      interval: [-0.1, 0.9]
      rejection_threshold: 200

pipelines:
  "TNN": 
    - name : TwinNeuralNetwork
      from: neuroIDBench.featureExtraction
      parameters: 
        EPOCHS: 10
        batch_size: 256
        verbose: 1
        workers: 1
\end{lstlisting}

\begin{lstlisting}[basicstyle=\ttfamily\small, caption={Benchamrking pipeline for dataset BrainInvaders15a with shallow classifiers and twin NN}]

name: "BrainInvaders2015a"

dataset: 
  - name: BrainInvaders2015a
    from: neuroIDBench.datasets
    parameters: 
      subjects: 10
      interval: [-0.1, 0.9]
      rejection_threshold: 200

pipelines:
    "TNN": 
    - name: TwinNeuralNetwork
      from: neuroIDBench.featureExtraction
      parameters: 
        EPOCHS: 10
        batch_size: 256
        verbose: 1
        workers: 1

    "AR+PSD+SVM": 
    - name: AutoRegressive
      from: neuroIDBench.featureExtraction
      parameters: 
        order: 6
        
    - name: PowerSpectralDensity
      from: neuroIDBench.featureExtraction
        
    - name: SVC
      from: sklearn.svm
      parameters: 
        kernel: 'rbf'
        class_weight: "balanced"
        probability: True
\end{lstlisting}

\begin{lstlisting}[basicstyle=\ttfamily\small, caption={Benchamrking pipeline for multi-session dataset COGBCI: FLANKER with shallow classifiers and Twin NN}]
name: "COGBCIFLANKER"

dataset: 
  - name: COGBCIFLANKER
    from: neuroIDBench.datasets
    parameters: 
      subjects: 10
      interval: [-0.1, 0.9]
      rejection_threshold: 200

pipelines:
    "TNN": 
    - name: TwinNeuralNetwork
      from: neuroIDBench.featureExtraction
      parameters: 
        EPOCHS: 10
        batch_size: 256
        verbose: 1
        workers: 1

    "AR+PSD+SVM": 
    - name: AutoRegressive
      from: neuroIDBench.featureExtraction
      parameters: 
        order: 6
        
    - name: PowerSpectralDensity
      from: neuroIDBench.featureExtraction
        
    - name: SVC
      from: sklearn.svm
      parameters: 
        kernel: 'rbf'
        class_weight: "balanced"
        probability: True
        
\end{lstlisting}

\begin{lstlisting}[basicstyle=\ttfamily\small, caption={Benchamrking pipeline for User i.e., Reseracher's own MNE data with shallow classifiers and Twin NN}]

name: "User"

dataset: 
  - name: UserDataset
    from: neuroIDBench.datasets
    parameters: 
      dataset_path: <local_dataset_path>
    
pipelines:
  "AR+PSD+SVM": 
    - name: AutoRegressive
      from: neuroIDBench.featureExtraction
      parameters: 
        order: 6
        
    - name: PowerSpectralDensity
      from: neuroIDBench.featureExtraction
        
    - name: SVC
      from: sklearn.svm
      parameters: 
        kernel: 'rbf'
        class_weight: "balanced"
        probability: True

  "TNN": 
    - name : TwinNeuralNetwork
      from: neuroIDBench.featureExtraction
      parameters: 
        EPOCHS: 10
        batch_size: 256
        verbose: 1
        workers: 1

  
    "AR+PSD+RF": 
        - name: AutoRegressive
        from: neuroIDBench.featureExtraction
        parameters: 
        order: 6
    
        - name: PowerSpectralDensity
        from: neuroIDBench.featureExtraction
      
        - name: RandomForestClassifier
        from: sklearn.ensemble
        parameters: 
            class_weight: "balanced"
        
\end{lstlisting}

\begin{lstlisting}[basicstyle=\ttfamily\small, caption={Benchamrking pipeline for User i.e., Reseracher's own EEG data and Researchers's own customized method for Twin NN}]

name: "User"

dataset: 
  - name: UserDataset
    from: neuroIDBench.datasets
    parameters: 
      dataset_path: <local_dataset_path>
    
pipelines:
  "TNN": 
    - name : TwinNeuralNetwork
      from: neuroIDBench.featureExtraction
      parameters: 
        user_tnn_path: <local_path_to_TNN>
        EPOCHS: 10
        batch_size: 256
        verbose: 1
        workers: 1
        
\end{lstlisting}

\subsection{Automated Benchmarking}
After setting the parameters for the benchmarking for single dataset. The automated benchmarking can be performed either by running the automation script from the command line terminal or by running the script within the Docker container. 
\\
\\
\textbf{Running the tool using command line terminal}
\\
\\ Execute the following command from the root directory:
\\
   python brainModels/run.py
\\
\\
\textbf{Running the tool using Docker}
\\
\\ 
Open the docker desktop installed on our local machine and execute the docker image my\_dockerfile.

\end{document}